\documentclass[prb,showpacs,superscriptaddress,twocolumn]{revtex4-1}
\pdfoutput=1
\usepackage[pdftex]{graphicx}  
\usepackage{graphicx, xcolor}
\usepackage{dcolumn}
\usepackage{bm}
\usepackage{amsmath,amsthm,amssymb}
\usepackage{color}
\usepackage{verbatim}

\definecolor{darkGreen}{RGB}{0,110,0}
\definecolor{darkBlue}{RGB}{0,0,130}
\usepackage[colorlinks,citecolor=darkGreen,linkcolor=darkBlue,urlcolor=blue,hyperindex]{hyperref}

\def\tit#1{}

\begin{document}
\title{Phenomenology of anomalous transport in disordered one-dimensional systems}

\author{M.\ Schulz}
\affiliation{The Abdus Salam International Center for Theoretical Physics, Strada Costiera 11, 34151 Trieste, Italy}
\author{S.R.\ Taylor}
\affiliation{The Abdus Salam International Center for Theoretical Physics, Strada Costiera 11, 34151 Trieste, Italy}
\author{A.\ Scardicchio}
\affiliation{The Abdus Salam International Center for Theoretical Physics, Strada Costiera 11, 34151 Trieste, Italy}
\affiliation{INFN, Sezione di Trieste, Via Valerio 2, 34126 Trieste, Italy}
\author{M.\ \v Znidari\v c}
\affiliation{Physics Department, Faculty of Mathematics and Physics, University of Ljubljana, Jadranska 19, SI-1000 Ljubljana, Slovenia}

\date{\today}

\begin{abstract}
We study anomalous transport arising in disordered one-dimensional spin chains, specifically focusing on the subdiffusive transport typically found in a phase preceding the many-body localization transition. Different types of transport can be distinguished by the scaling of the average resistance with the system's length. We address the following question: what is the distribution of resistance over different disorder realizations, and how does it differ between transport types? In particular, an often evoked so-called Griffiths picture, that aims to explain slow transport as being due to rare regions of high disorder, would predict that the diverging resistivity is due to fat power-law tails in the resistance distribution. Studying many-particle systems with and without interactions we do not find any clear signs of fat tails. The data is compatible with distributions that decay faster than any power law required by the fat tails scenario. Among the distributions compatible with the data, a simple additivity argument suggests a Gaussian distribution for a fractional power of the resistance.
\end{abstract}

\maketitle

\section{Introduction}

The steady-state transport of a globally conserved quantity is one of the simplest manifestations of non-equilibrium physics. Understanding transport properties of common quantum toy models is therefore of obvious theoretical importance, even more so is the applied component of the question. Furthermore, if the transport is slow, the relaxation of an initial non-equilibrium state to equilibrium will also be slow. Relaxation and thermalization properties~\cite{d2016quantum} are therefore intimately related to transport.

In one-dimensional (1D) systems transport properties can be especially rich. On the one hand there is a possibility to have integrable systems for which, at least if they are translationally invariant, one generically expects ballistic transport. One can understand that through the existence of nontrivial conservation laws~\cite{Zotos1997Transport}, or behavior of appropriate elementary excitations. Be it in an interacting or a free model, they propagate without dissipation resulting in a ballistic scaling of resistance $R$ with system length $L$  as $R \sim L^0$, i.e. the resistance does not increase with $L$. The other extreme situation is that of localization, an example being Anderson localization~\cite{Anderson1958} (i.e. non-interacting particles), for which $R$ is exponentially large, $R \sim \exp{(L/\xi)}$. In 1D non-interacting particles localize for any strength of an on-site potential. More recently it has been realized~\cite{Basko:2006hh} that interactions do not necessarily cause a breakdown of localization and that many-body localization (MBL) is actually possible~\cite{gornyi2005interacting,huse2015review,abanin2017recent,imbrie2017review,Alet18,Abanin2019colloqium}, although the situation is more complicated in geometries with higher spatial dimensions.~\cite{chandran2016many,de2017stability} In a nutshell, 1D interacting systems can present ballistic transport in the absence of disorder, and are completely localized for sufficiently large disorder.

A natural and not yet fully understood question concerns disorder strengths intermediate between the clean and the localized extremes. Does the transport type vary continuously, or are there phase transitions, and if yes, how and why do they occur? From the point of view of (quantum) chaos theory~\cite{dorfman99} one could argue that the disorder, which renders models quantum chaotic, should in general result in diffusive transport for which normal (Ohm's) scaling holds, $R \sim L$. However, perhaps surprisingly, this is not always the case. 

In the Heisenberg model with a random on-site magnetic field (which is equivalent to interacting spinless fermions on a lattice, with on-site random potentials) it has been observed that the transport can in fact be subdiffusive~\cite{Reichman2014Absence,Gopalakrishnan2015low,torresherrera2015dynamics,Khait2016spin,Luitz2016Extended,prelovchek2017self,doggen2018many,AgarwalAnomalousDiffusion,Znidaric2016Diffusive,Varma2017Energy,mendoza2018asymmetry,schulz2018energy,Karahalios2009finite,Steinigeweg2016typicality,Roek2019prep} with $R \sim L^\gamma$, with $\gamma>1$ and that there is a phase transition at finite disorder strength $W_{\rm c}$ from diffusive to subdiffusive transport~\cite{AgarwalAnomalousDiffusion,Znidaric2016Diffusive,Varma2017Energy,mendoza2018asymmetry,schulz2018energy,Karahalios2009finite,Steinigeweg2016typicality}. Subdiffusion has also been related to anomalous distributions of eigenstate matrix elements~\cite{Luitz2016Anomalous}.

An explanation for the subdiffusive transport that was offered is the so-called Griffiths effect picture~\cite{Griffiths, Gopalakrishnan2016Griffiths,Agarwal2017rare,Potter2015Universal}. This ascribes the slow dynamics to rare blocking regions of locally higher disorder that are bound to greatly influence transport in 1D systems. One of the predictions of the Griffiths picture is fat tails~\cite{AgarwalAnomalousDiffusion,AltmanTheory2015,Gopalakrishnan2017Noise} in the distribution $p(R)$, $p(R) \asymp 1/R^\beta$, such that the average $R$ does not exist and therefore the law of large numbers does not apply, and the total resistance of a sample of size $L$ scales super-linearly with $L$.

While this could explain the dependence of the typical $R(L)$, so could other scenarios. Beyond the scaling exponents for average quantities, the Griffiths picture has not undergone any independent test so far. It is therefore important to check other quantities to really see if the Griffiths scenario is the correct phenomenological description of subdiffusion. 
Another reason that it is important to better understand the effects of rare regions is because they feature prominently in many proposals to describe the MBL transition using renormalization group techniques~\cite{Vosk:2013kq,PhysRevB.90.224203,AltmanTheory2015,Potter2015Universal}. 
In those theories, the MBL transition arises due to competition between conducting and (rare) insulating regions and as such the properties of rare regions do matter.

There is also an alternative scenario for $p(R)$ in which $R\sim L^\gamma$ but its distribution does not have fat tails. Namely, if we assume that the distribution of $R^{1/\gamma}$ has at least two moments, then the law of large numbers should lead to a Gaussian distribution of a quantity that is extensive (i.e., additive in $L$). Such arguments have in fact been used a long time ago~\cite{Anderson1980new} when looking for the correct scaling variable of Anderson localization. This ``additivity argument" results in general anomalous transport with the scaling $R \sim L^\gamma$ in the thermodynamic limit (TDL), while the variable $x:=R^{1/\gamma}$ is normally (Gaussian) distributed, so no fat tails are present in $p(R)$.

A microscopic mechanism for subdiffusion alternative to fat tails and rare regions is also indicated by the Fibonacci model, both the non-interacting~\cite{Hisashi1988dynamics,piechon1996anomalous} and the interacting one~\cite{Mace2019many-body,Varma2019Diffusive}, where the potential is deterministic and quasiperiodic, and therefore strictly speaking there are no rare regions, but subdiffusion is still observed.

While a microscopic understanding of transport would be the preferred goal, in the absence of any existing analytical picture, in this work we aim for a more modest goal of first understanding the phenomenology of different transport types, in particular the distribution of resistance $p(R)$. 
In non-interacting systems, where we can probe large system sizes, distributions are always compatible with the above-mentioned additivity, that is $p(x=R^{1/\gamma})$ is Gaussian. For the interacting disordered Heisenberg model numerics are more difficult and only smaller systems are available, however, we can say that for system sizes of order $L \sim 50$, if we insist on fitting to a power-law tailed distribution, the best fit power is large and, furthermore, it increases with $L$. 
A plausible scenario is therefore that in the TDL the additivity of $R^{1/\gamma}$ would again result in a Gaussian $p(R^{1/\gamma})$. We therefore conclude that the fat tails scenario of the Griffiths picture cannot fully describe the distribution of resistance in the subdiffusive regime, at least not on the scales available to present-day numerics and far away from the MBL transition.

\section{Model and description of method}
\label{sec:method}

We will henceforth consider the XXZ model typically studied~\cite{vznidarivc2008many} in the MBL context:
\begin{equation}
    H=\sum_{i=1}^L s^{\rm x}_i s^{\rm x}_{i+1} + s^{\rm y}_i s^{\rm y}_{i+1} + \Delta s^{\rm z}_i s^{\rm z}_{i+1} + h_i s^z_i
    \label{equ:Heisenberg}
\end{equation}
where $s^{\alpha}_i = \frac{1}{2}\sigma^{\alpha}_i$ are spin-$1/2$ operators ($\sigma^{\alpha}_i$ are Pauli matrices), and $h_i$ are random fields at site $i$ which are i.i.d. numbers drawn from a distribution specified when relevant. In this paper we look at either the non-interacting model where $\Delta=0$, usually called the $XX$ model, or the interacting XXX Heisenberg model with $\Delta=1$. For the interacting case, as soon as $h>0$, at infinite temperature, a diffusive region exists, giving way to a subdiffusive region starting at $W \sim 0.5$~\cite{Znidaric2016Diffusive} (for a box distribution of width $2W$), i.e. relatively quickly after breaking integrability and rather far from the proposed MBL transition\cite{pal2010many, alet2015} $W_{\text{MBL}}\sim 4$.

To be as close to the TDL as possible we use an open system formulation in which the XXZ chain is driven at the boundaries by two `baths', such that after a long time the system ends in a non-equilibrium steady-state (NESS). The expectation value of the spin current $j_k \equiv s^x_k s^y_{k+1} - s^y_k s^x_{k+1}$ in the NESS is then our main observable whose distribution we study. In order to simulate the open system we employ a super-operator version of the time-evolving block decimation (TEBD) method~\cite{schollwock2011density} used in similar studies \cite{Znidaric2016Diffusive,Mendoza2015Coexistence,mendoza2018asymmetry}. Concretely, we simulate a boundary-driven Lindblad master equation \cite{gorini1976completely,lindblad1976generators,Breuer2002} describing Markovian time-evolution of a system density matrix:
\begin{equation}
\frac{d\rho}{dt}\ = i\left[\rho,H \right] +\frac{1}{4} \kappa \sum_{k=1}^{4} \left(\left[L_k \rho,L_k^{\dagger}\right] + \left[L_k, \rho L_k^{\dagger}\right]  \right),
\label{equ:lindblad}
\end{equation}
where $H$ describes the closed system disordered Heisenberg chain \eqref{equ:Heisenberg}, and where the Lindblad operators $L_k$ account for generic magnetization driving by two `baths'. They are defined as $L_1 = \sqrt{1+\mu}\sigma^+_1$ and $L_2 = \sqrt{1-\mu}\sigma^-_1$ on the left side, and $L_3 = \sqrt{1+\mu}\sigma^-_L$ and $L_4 = \sqrt{1-\mu}\sigma^+_L$ on the right side. Unless mentioned otherwise, we use $\kappa=1$. With $\mu \neq 0$ (we use $\mu=0.001$ in this paper), there is an asymmetry in the driving between the two sides and a non-zero, unique NESS magnetization current is induced. As previously discussed in \cite{Znidaric2016Diffusive}, a microscopic derivation of such a drive might be difficult, however in a generic non-integrable model the details of the boundary drive should not matter for the bulk physics.

For a given initial state we time-evolve until a series of convergence criteria are fulfilled. The central quantity of this simulation is the magnetization or spin current. Its expectation value $\text{Tr}\left(j_k \rho_{\infty}\right) \equiv j$ is, due to stationarity, independent of the site index $k$. We use this spatial homogeneity of the spin current as the most suitable indicator of convergence. Further we also demand temporal uniformity\footnote{The temporal criterion is that the standard deviation of the average current over the previous 100 time-steps is less than $0.3\%$ of the average over the same period.} of the spin current in order to exclude a slow but spatially uniform drift in the current. We choose as our spatial criterion that the standard deviation of the individual currents on every bond $k$ (excluding those subjected directly to the Lindblad driving) relative to the average current be $\sigma(j_k)/j \sim 2 \%$ (at most $4 \%$ for the largest disorder, where convergence is slower).

In the numerical TEBD algorithm, one fixes a bond dimension $\chi$ for the representation of the state $\rho$ and tries to find an approximate NESS with this form of the state. The higher $\chi,$ the better the approximation to the true NESS. We start the search with a relatively low $\chi$ and we take $\rho(0)$ for each disorder realization as a completely mixed local density matrix (infinite temperature). We then use the obtained NESS at fixed $\chi$ as the initial state for a simulation with a higher matrix dimension $\chi$, up to a maximum of $\chi=800$.  As the NESS is unique, we make the reasonable assumption that every increase of $\chi$ will bring our numerical approximation of the current $j$ closer to its true value. This proves more difficult for stronger disorder and is discussed later. We present a possible extrapolation that may be performed so that we may approximate the fully converged current $j(\chi \rightarrow \infty)$.

For the discrete time-steps performed in TEBD, we employ a fourth-order Trotter decomposition, mostly with a time-step of $dt = 0.4$ for local two-site updates. We checked that the results do not change with decreasing $dt$ within the tolerance described before.

Finally, we should mention that in all the cases studied so far, where a careful comparison has been made, the Lindblad setting gives exactly the same transport as for instance linear response calculations based on Green-Kubo type formulas; for diffusion one can show this analytically~\cite{znidarivc2019nonequilibrium}, while for subdiffusion it has been verified numerically~\cite{Varma2019Diffusive}. The Lindblad equation is therefore just a tool that enables us to study systems an order of magnitude larger than would be possible with other methods.

\section{Non-interacting models}

\begin{figure}
    \centering
    \includegraphics[width=.45\textwidth]{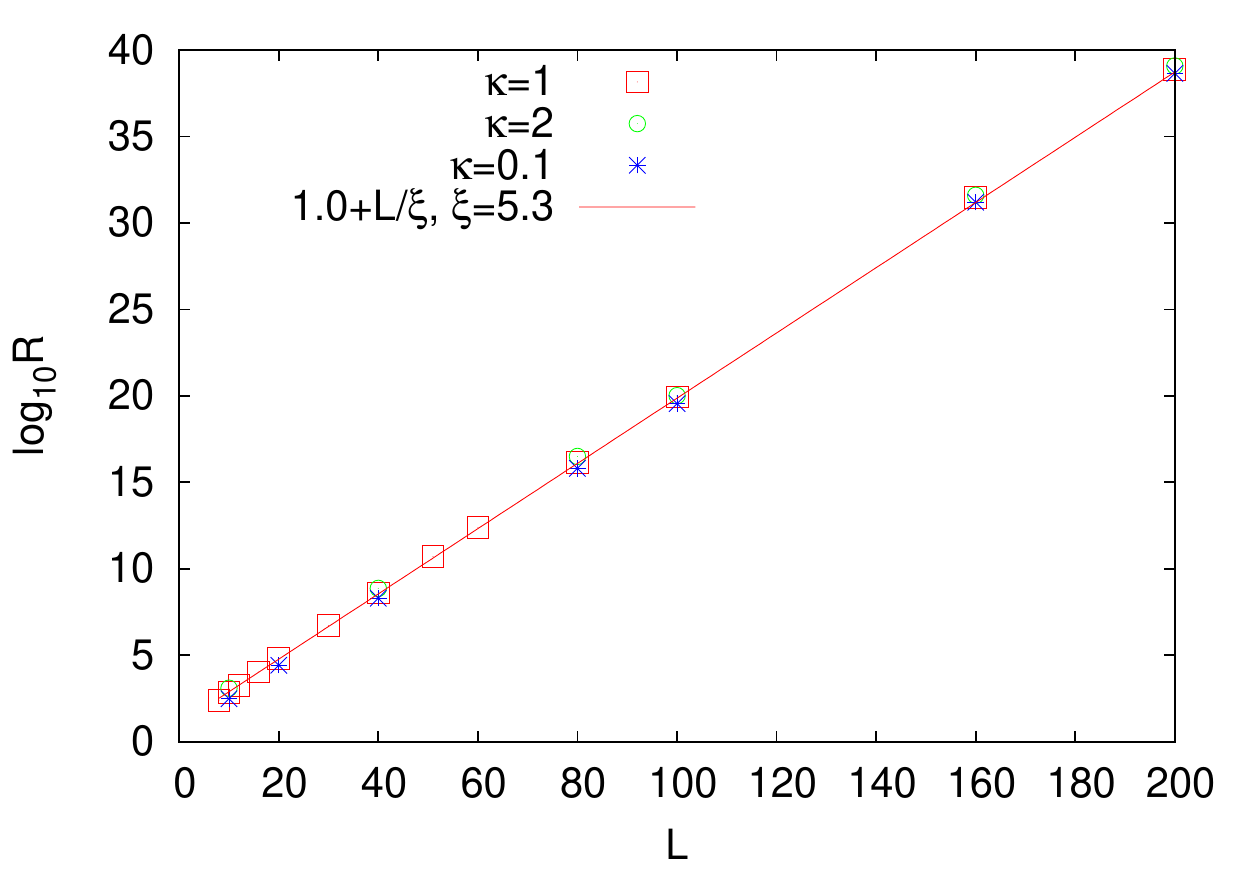}\\
    \caption{Scaling of the average $\log_{10}{R}$ with system length $L$ for the Lindblad-driven Anderson model with disorder strength $W=2$, and different coupling strength $\kappa$ in the Lindblad equation.}
    \label{fig:RodL}
\end{figure}

We shall first check the distribution $p(R)$ for non-interacting models, where numerics are easier and one can use correspondingly larger $L$. An added advantage is that we will also gain insight into how large $L$ has to be in order for $p(R)$ to converge to its TDL.

We are going to check $p(R)$ in two non-interacting models, one will be non-interacting particles with an on-site disorder (the Anderson model, or, equivalently, the XX spin chain with disorder) that shows localization, while the other will be the Fibonacci model in a regime with subdiffusion. The Hamiltonian for both is \eqref{equ:Heisenberg} with $\Delta=0$, and we use the boundary-driven Lindblad formalism to get the NESS and the associated resistance $R$ for a particular potential realization $h_i$ as described in Sec.~\ref{sec:method}. For the Anderson model the goal is to check that indeed, also in the Lindblad setting we get the expected result known from a closed Hamiltonian formulation, namely that $R$ is exponentially large in $L$ and log-normally distributed~\cite{Anderson1980new,abrikosov1981paradox}. This should disperse any doubts that the Lindblad setting could somehow crucially influence the results we are presenting.

\begin{figure}
    \includegraphics[width=.45\textwidth]{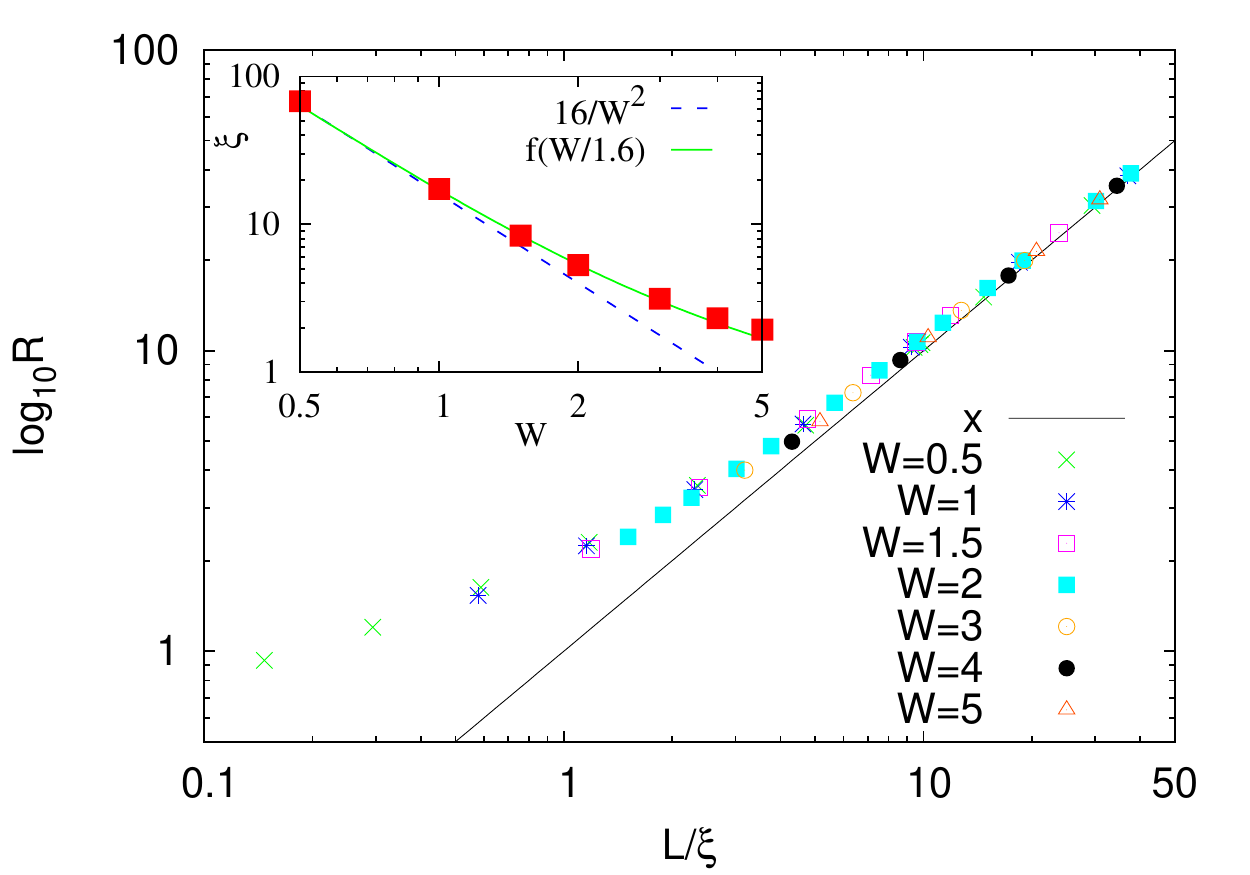}
    \caption{Single-parameter scaling of average $\log_{10}{R}$ for Anderson localization. Data is shown for various $W$ and $L$, for instance, for $W=5$ sizes up to $L=60$ are shown, while for $W=0.5$ up to $L=2000$. Inset: for small disorder the localization length is $\xi \sim 1/W^2$, while for larger disorder a better description is obtained by the scaling function $f(x):=1/(-1+\frac{1}{2}\ln[1+x^2]+\frac{1}{x}\arctan{x})$, green curve, see Ref.~\onlinecite{izrailev1998classical}.}
    \label{fig:scaling}
\end{figure}

\subsection{Additivity}
\label{sec:additivity}
Let us first recall the additivity argument~\cite{Anderson1980new} as it will turn out that in both non-interacting models $p(R)$ agrees with its predictions. For an arbitrary transport with scaling $R \sim L^\gamma$, one can argue that a quantity which is additive will converge to a Gaussian distribution in the TDL. This gives the correct result in the two limiting cases: (i) localization, and (ii) diffusion. Namely, for diffusion one does not expect any complications, leading to a Gaussian distribution of $R$ -- later on we will check that this is indeed the case for the interacting disordered Heisenberg model in the diffusive phase. For localization, where one has $R \sim \exp{(L/\xi)}$, with $\xi$ being the localization length, one therefore expects that $\lim_{a \to 0} R^a$ should be additive, leading naturally to the logarithm via $\log{x}=\lim_{a \to 0}(x^a-1)/a$. And indeed, the logarithm of $R$, which is equal to $L/\xi$, is extensive and is normally distributed~\cite{Anderson1980new}. Based on the results obtained in the present paper we conjecture that the additivity may give the correct prediction for $p(R)$ for any transport type.

\subsection{Anderson localized system}

For systems of non-interacting fermions, e.g. the $XX$ spin chain, and for our choice of Lindblad operators the whole Liouvillian (linear operator corresponding to the RHS of Eq.~\ref{equ:lindblad}) is quadratic in fermonic variables and can therefore be efficiently numerically solved because one needs to diagonalize only a matrix of size~\cite{prosen2008third} $\sim L$ instead of a full Liouvillian which is of size $4^L$.

To probe Anderson localization in a Lindblad setting we therefore take the $XX$ spin chain, Eq. \eqref{equ:Heisenberg} with $\Delta = 0$, with on-site disorder $h_i$, where $h_i$ are i.i.d. random variables with Gaussian distribution with zero mean and standard deviation\footnote{The odd-looking $\sqrt{3}$ is here in order to have the same variance $W^2/3$ as one would have for the often-used box distribution $h_j \in [-W,W]$.} $W/\sqrt{3}$.

\begin{figure}
    \centering
    \includegraphics[width=\columnwidth]{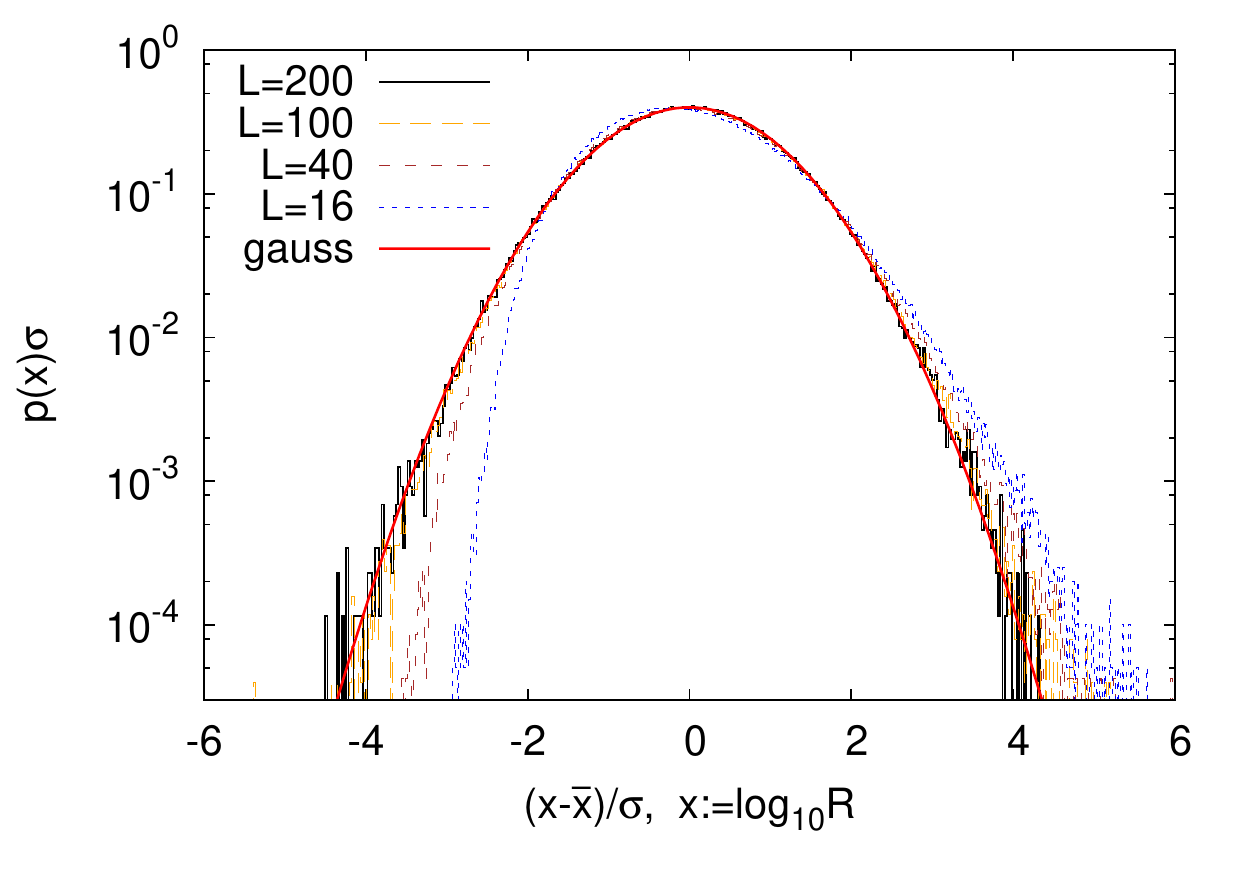}
    \caption{Distribution of the scaled logarithmic resistance for the Anderson model with Gaussian disorder of strength $W=2$. Variance is $\sigma^2:=\langle x^2\rangle-{\langle x \rangle}^2$. For sufficiently large $L$ the distribution of $\log{R}$ is Gaussian.}
    \label{fig:anders_dist}
\end{figure}

We generate many instances of disorder realization, calculating for each the steady-state and in particular its NESS current $j_\infty$, see e.g. Ref.~\cite{znidarivc2013transport} for technical details. Because the system is quadratic, $j_\infty$ is trivially and exactly proportional to the driving parameter $\mu$. The resistance is then simply
\begin{equation}
  R=\mu/j_\infty.
  \label{eq:R}
\end{equation}
We first check that the resistance is indeed exponentially large in $L$. This is shown in Fig.~\ref{fig:RodL}. The coefficient of the linear slope of $\log R \sim L / \xi$ is nothing but the localization length $\xi$. We can also see that asymptotically $R$ does not depend on the coupling strength $\kappa$ used in the Lindblad equation \eqref{equ:lindblad}.
Next we check predictions of the scaling theory which says that $R$ is not an independent function of $L$ and $W$, but rather a function of a single scaling variable $L/\xi$. This is shown in Fig.~\ref{fig:scaling}.

Finally, we show the distribution of the resistance, our main object of study. Gathering $10^6$ disorder realizations (except for $L=200$ where we have $3\cdot 10^5$) we plot histograms of the distribution in Fig.~\ref{fig:anders_dist}. One can see that a fairly large size $L \approx 100$ is required in order to converge to a Gaussian. At smaller sizes, for instance $L=16$, the distribution is skewed towards larger $R$ -- there are too many large-$R$ instances. We also remark that getting statistics for large $L$ is harder not only because one has to deal with larger matrices but also because larger precision is required. For instance, for $L=16$ one has $\overline{\log_{10}{R}}\approx 4.0$ and the standard deviation of the distribution is $\sigma \approx 1.0$, while for $L=200$ one has $\overline{\log_{10}{R}}\approx 39$ and $\sigma \approx 3.9$ (to that end $j_\infty$ had to be calculated to 60 digits of precision because the calculations become unstable once the current gets smaller than the standard floating-point machine precision $\sim 10^{-16}$. To remedy that one has to increase the number of digits of precision, which can for instance be done easily in Mathematica).

Concluding this Anderson part we can say that the results obtained are all expected -- the main goal was to check that the Lindblad setting does not introduce any spurious effects. Furthermore we see that fairly large systems are required in order to converge to the TDL distribution $p(R)$.

\subsection{Subdiffusive non-interacting system}
\begin{figure}
    \centering
    \includegraphics[width=\columnwidth]{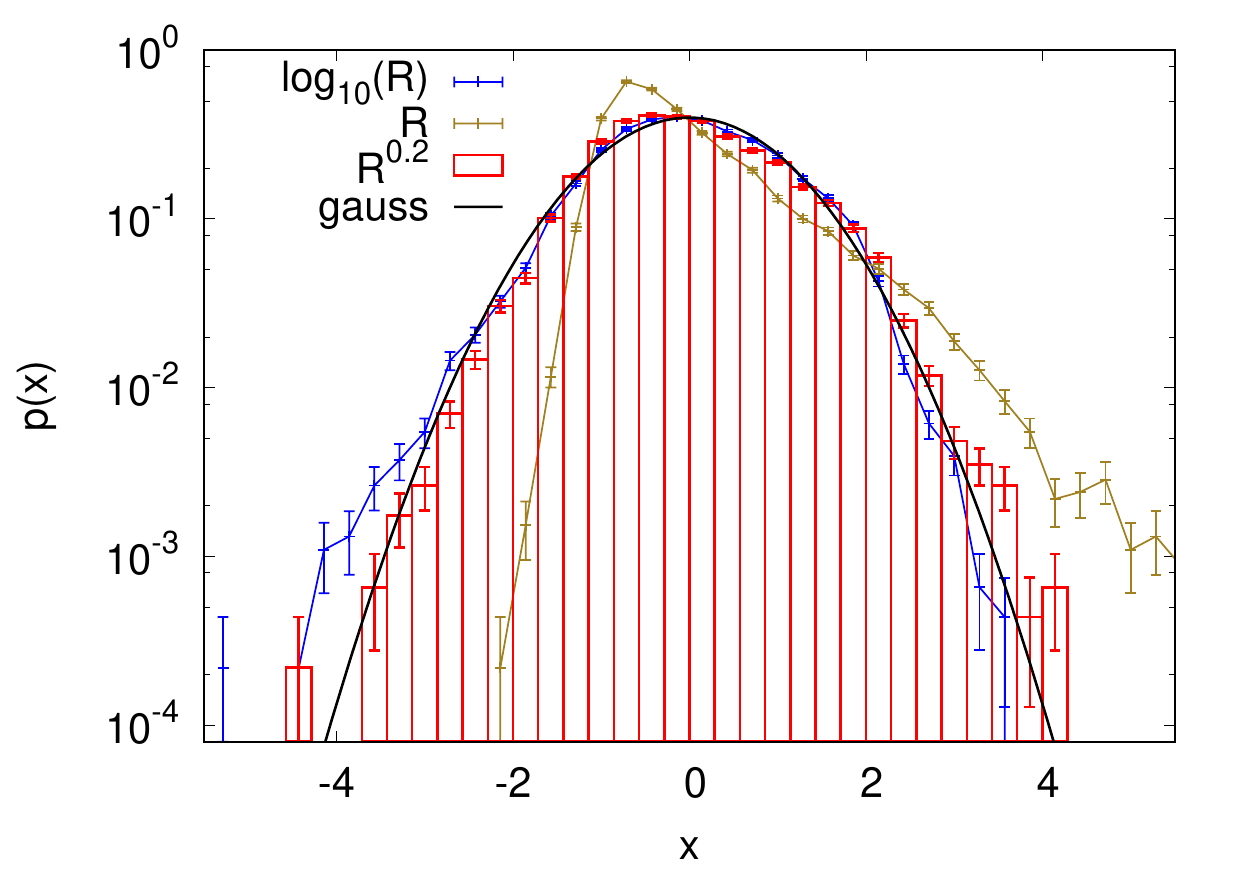}
    \caption{Distribution $p(x)$ for the non-interacting Fibonacci model with $W=1.5$ and three different choices of $x$: resistance $R$ (olive), the logarithm $\log_{10}{R}$ (blue), and $R^{\alpha=0.2}$ (red). Black curve is a Gaussian $\frac{1}{\sqrt{2\pi}}{\rm e}^{-x^2/2}$. In all cases we show variables that have mean zero and variance $1$, for instance, in the $R$-case $x:=(R-\overline{R})/\sigma(R)$.}
    \label{fig:fiboP}
\end{figure}

The Fibonacci model~\cite{kohmoto1983localization,ostlund1983one} is a model of non-interacting particles (XX spin chain) in which an on site potential $h_j$ has strength $h_j=W$ or $h_j=-W$, depending on the site index $j$. There are several equivalent ways of how to specify a quasiperiodic Fibonacci pattern of disorder. A compact way is writing $h_j=W (2V(jg)-1)$, with an irrational $g=(\sqrt{5}-1)/2$ and periodic $V(x):=[x+g]-[x]$ with $[x]$ being an integer part of $x$. One of the interesting properties of the Fibonacci model is that its transport type, i.e. the scaling exponent $\gamma$, continously varies~\cite{Hisashi1988dynamics,piechon1996anomalous} with potential amplitude $W$ from ballistic $\gamma=0$ at $W=0$ all the way to localized $\gamma \to \infty$ in the limit $W \to \infty$. In particular, for sufficiently large $W$ ($W$ larger than about $0.8$) one has subdiffusion.  

In the following we pick $W=1.5$, a Lindblad coupling strength $\kappa=1$, and study the NESS current $j_\infty$ and the associated resistance $R$ (\ref{eq:R}). In order to study the distribution $p(R)$ we need an appropriate ensemble. Because $H$ is fully deterministic for the Fibonacci model, i.e. there is no explicit disorder, the best one can do is to take a system of length $L$ starting with potential $h_j$ at a given site $j_0$ different from the 1st one. In a system of length $L$ one can get $L+1$ different $H$ in such a way, see e.g. Ref.~\onlinecite{Mace2019many-body}. Due to reflection symmetry one however gets only $\sim L/2$ different NESS currents. Our ensemble therefore consists of about $L/2$ instances for a given $L$. The Liouvillian is again quadratic so systems with $L \sim 10^3$ can be studied. Still, the ensemble size $L/2$ is too small for the purposes of obtaining the distribution. To that end we take $L \approx 1000$ as well as several neighboring $L$ and average data over all those $L$.

The scaling exponent of $R$ is about $\gamma \approx 2.0$ (which is slightly different from $\gamma=1/\beta-1\approx 1.6$ obtained from a wave-packet spreading in Ref.~\onlinecite{Varma2019Diffusive}), however one should be aware that possible different scaling exponents could emerge using different sequences of $L$ (see the Aubry-Andre case for an example~\cite{Varma2017fractality}). The additivity argument would therefore predict that $x:=R^{\alpha}$ with $\alpha=1/\gamma \approx 0.5$ should be Gaussian distributed. However, on plotting $p(x)$ we noticed that we get better agreement with a Gaussian upon taking the exponent $\alpha$ to be smaller than $0.5$, the best fit being obtained for $\alpha \approx 0.2$. 

\begin{figure}
    \centering
    \includegraphics[width=\columnwidth]{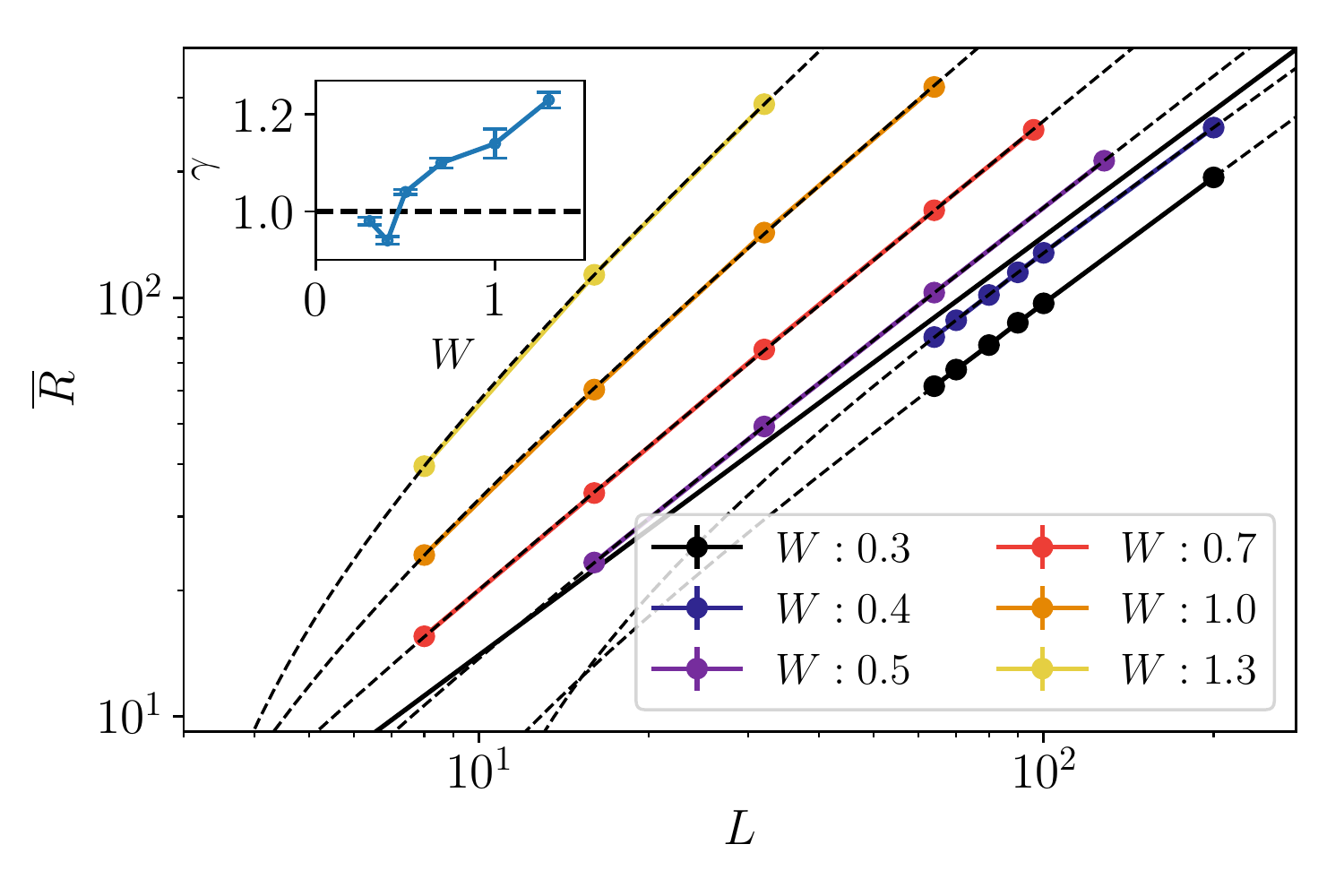}
    \caption{Scaling analysis of the average current $\overline{R}$ fitted to $R = c_0 L^{\gamma} (1 + c_1/L)$ for the XXX model. The black dashed lines indicate a fit to the data and the black solid line denotes a slope of unity, i.e. diffusion. The inset shows the values of $\gamma$ obtained as a function of $W$, where the horizontal dashed line denotes a diffusive $\gamma = 1$.}
    \label{fig:R_L}
\end{figure}

This $\alpha$ is then used in Fig.~\ref{fig:fiboP} where we plot a histogram of all data for a range of system sizes $L=978-1009$, resulting in a total 15,965 independent values of $R$. We plot the distribution of three different quantities. We can see that $p(R)$ (olive) is clearly not Gaussian -- there are far too many large $R$ instances. The distribution of $\log_{10}R$ (blue), being the limit of $\alpha \to 0$, differs less from a Gaussian distribution (black). While it might be difficult to distinguish the two distributions, we do however have a statistically significant difference (a couple of sigma) in a number of consecutive bins at the left edge (small negative $x$). The distribution of $R^{0.2}$ (red) fits within statistical error to a Gaussian (black), and therefore seems compatible with the additivity prediction that the distribution of $R^\alpha$ is Gaussian.

At present it is not clear why the best exponent $\alpha \approx 0.2$ differs from $1/\gamma$. In fact, there can be several reasons for this, among them being multifractality of the model as well as finite $L$ (and merging statistics for different $L$). Nevertheless, we can certainly say that in this, undoubtedly subdiffusive, model there are no fat power-law tails in the distribution of $R$. This could be checked for fairly large systems $L \sim 1000$, much larger than is possible for the XXX model which we focus on next.

\section{Interacting disordered Heisenberg model}

\begin{figure}
    \centering
    \includegraphics[width=\columnwidth]{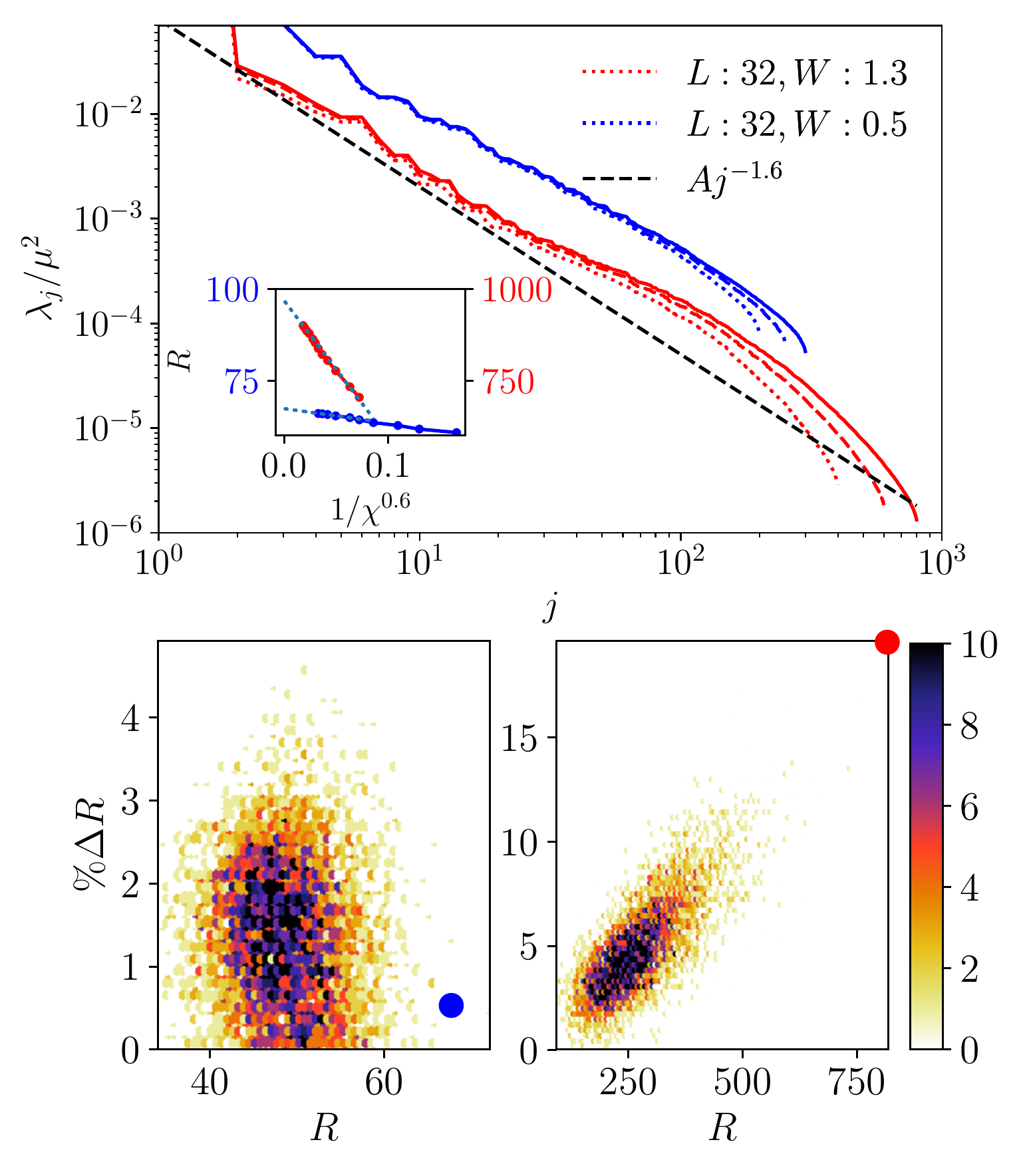}
    \caption{The top panel shows the Schmidt values of the middle bond of the MPO describing the NESS for two instances of disorder for $L = 32$ and $W = 0.5,1.3$. The different line-styles denote simulations with different maximum bond-dimension $\chi_{\text{max}}$ ($\chi_{\text{max}}=400,600,800$ for $W=1.3$ and $\chi_{\text{max}}=200,250,300$ for $W=0.5$). The dashed black line indicates the power used in the resistance extrapolation $R(\chi\rightarrow\infty)$ as $1/\chi^{p-1}$ with $p=1.6$
    (shown in the inset for the same two instances).
    Bottom panels: We show a scatter plot of the percentage deviation of the extrapolated values of the resistance from the highest simulated $\chi=200$. The colored points indicate the instances taken for the top panel.}
    \label{fig:R_t_and_R_chi}
\end{figure}

In this section we study the XXX model, Eq.(\ref{equ:Heisenberg}) with $\Delta=1$, with a box distribution of disorder, $h_j \in [-W,W]$. As a test case for the open system simulation with interactions, we first aim to reproduce an appropriate estimate for the diffusion-subdiffusion transition as presented in \cite{Znidaric2016Diffusive}. In contrast to the previous section, the whole Liouvillian is no longer quadratic so we have to resort to a matrix-product operator (MPO)-based method. This method to obtain the NESS current is presented in Sec.~\ref{sec:method}. In Fig.~\ref{fig:R_L}, we perform the same analysis as Ref.~\onlinecite{Znidaric2016Diffusive}, in that we determine the NESS resistance $R$ for many disorder instances and system sizes $L$ and fit the resulting averaged values of $R$ to obtain a transport coefficient $\gamma$. We obtain a similar diffusion-subdiffusion transition around the disorder strength of $\sim W=0.5$. To estimate finite-size effects we employ a fit of the form $R=c_0 L^\gamma(1+c_1/L)$, which turns out to have the best chi-square per d.o.f.\ of any form we tried (in particular for $W=0.7$ it has half the $\chi^2$ of the form $R=c_0 L(1+c_1/L+c_2/L^2)$ which has the same number of d.o.f.'s). We observe that for $W=0.3, 0.4$, the best-fit $\gamma$ is consistent with 1 (at $90\%$ CL we have respectively $\gamma\in[0.97,1.01]$ and $[0.94,0.99]$), while for $W=0.5$, $\gamma=1.038\pm 0.005$ (with the value $1$ excluded at $90\%$ CL). For $W=0.7,\ \gamma=1.10\pm0.01$ and for $W=1.0,\ \gamma=1.14\pm0.03$. While the finite-size corrections become a little bit larger as $W$ grows ($c_1$ remains below $5$), the quality of the fit decreases (as indicated by considerably larger chi-square values), essentially because of the systematic errors due to the truncation of the bond dimension in our algorithm. For the largest $W=1.3$  the errors associated with the finite bond dimension truncation are larger than the statistical ones (which are very small, thanks to sample sizes of $3,000-40,000$). The precise values of $\gamma$ are not very reliable but the power-law fit remains the best fit. Moreover, the error due to truncation of the bond dimension is \emph{systematically underestimating} the resistance for larger $L,W$ and therefore underestimating $\gamma$ (and its error). 

In fact, during the search of the NESS, we observed that for some instances we do not reach convergence to the true NESS, due to the restriction of a finite $\chi$. This manifests itself in systematic upward drift of the value of the resistance when increasing the bond-dimension further, especially in the case of large disorder and therefore large resistance. In Fig.~\ref{fig:R_t_and_R_chi} we attempt to summarize the situation for large resistance instances. 
The key insight is that we may obtain a reasonable estimate of the error from the Schmidt spectrum of the MPO describing the density matrix $\rho$. In practice we take the non-zero Schmidt coefficients  for a bipartite splitting of a super-ket $\vert \rho \rangle$ into two halves of length $L/2$, denoted by $A$ and $B$, $\vert \rho \rangle = \sum_j \sqrt{\lambda_j} \vert \xi^{A}_{j} \rangle \vert \xi^{B}_{j} \rangle$, and integrate the spectrum to obtain the rate at which the weight of neglected Schmidt values decays to zero. We therefore make the reasonable conjecture that the correct NESS value of $R$ is reached in a similar way. If the Schmidt coefficients decay algebraically as $\lambda_j \sim 1/j^{p}$, then we expect $\sum_j \lambda_{\text{negl.}}\sim 1/\chi_{\text{max}}^{p-1}$.

The top panel in Fig.~\ref{fig:R_t_and_R_chi} reflects this analysis with the inset showing the extrapolation for two chosen instances.  We note that the spectrum does not change significantly for different disorder strengths and appears reasonably well converged with $\chi_\text{max}$ for the largest Schmidt values. While the extrapolation in $1/\chi^{p-1}$ works well for both instances, the prefactor is largely different between the small and large disorder instance, resulting in very different convergence properties of the NESS. We note that all figures concerning the interacting model use extrapolated values of the resistance $R(\chi\rightarrow\infty)$, obtained by using $1/\chi^{p-1}$ with $p=1.6$.
We thus display in the bottom panel of Fig.~\ref{fig:R_t_and_R_chi} an overview of the percentage deviation of the extrapolated values of the resistance from the highest simulated $\chi$, i.e. $\%\Delta R \equiv \vert (R(\chi\rightarrow\infty)-R(\chi_\text{max}))\vert/R(\chi_\text{max})$. The results agree with our intuition that at large disorder the NESS is increasingly difficult to obtain for finite $\chi$. 
Furthermore, in the case of large $W$ there is a positive correlation between the size of $R$ of the simulated instance and its associated extrapolation error. Extreme cases such as displayed in Fig.~\ref{fig:R_t_and_R_chi} are however rare and do not influence the probability distributions $p(R)$ significantly. The error bars in Fig.~\ref{fig:R_L} and for the obtained values of $\gamma$ contain an estimated error due to $\%\Delta R$ as well finite size effects $\sim A/L$.

\begin{figure}
    \centering
    \includegraphics[width=\columnwidth]{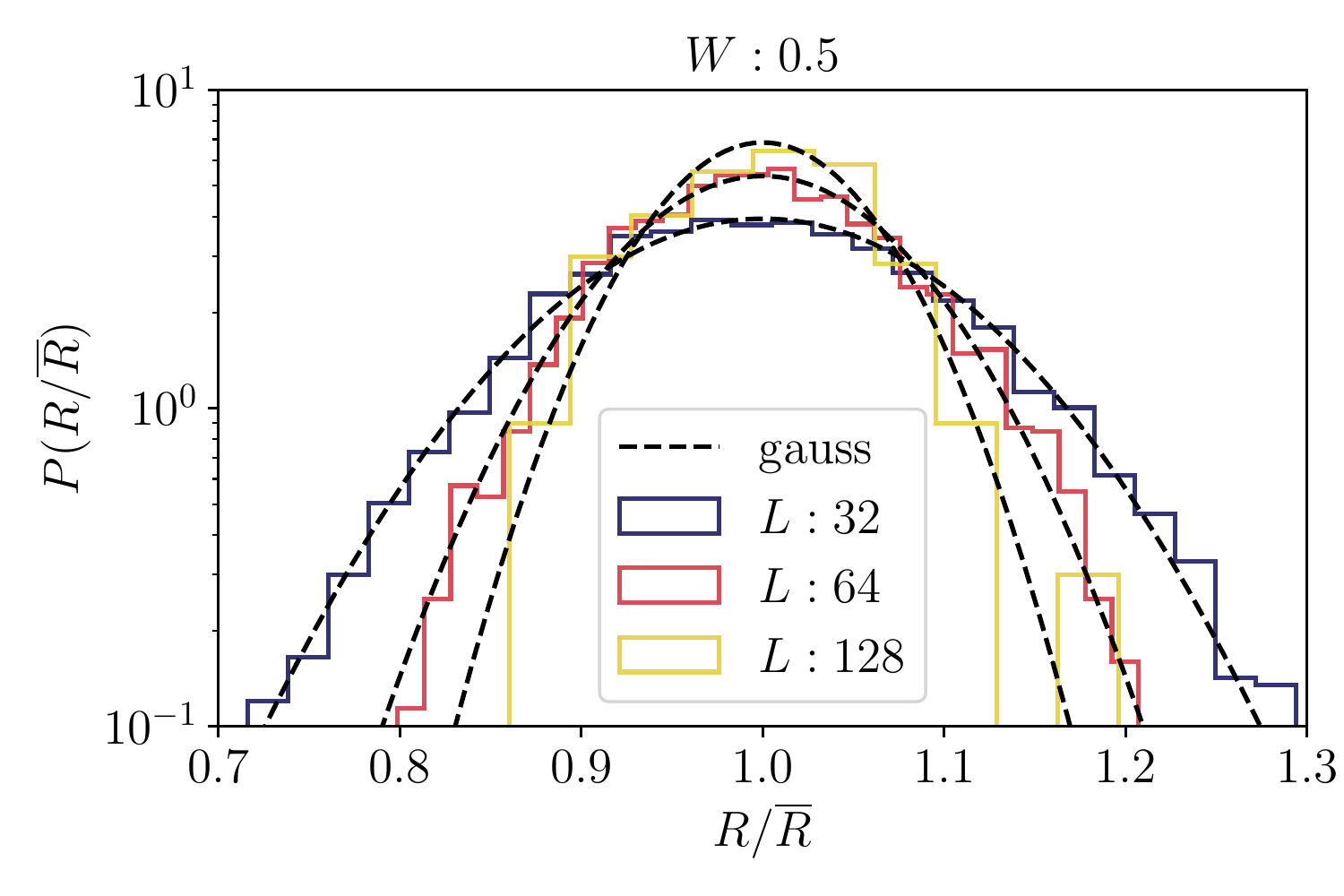}
    \caption{The probability distribution of the scaled resistance $R/\overline{R}$ for different system sizes at a disorder strength of $W = 0.5$ (roughly at the diffusion-subdiffusion transition). The black dashed line denotes a Gaussian fit to the data.}
    \label{fig:R_R_diff}
\end{figure}

\begin{figure}
    \centering
    \includegraphics[width=\columnwidth]{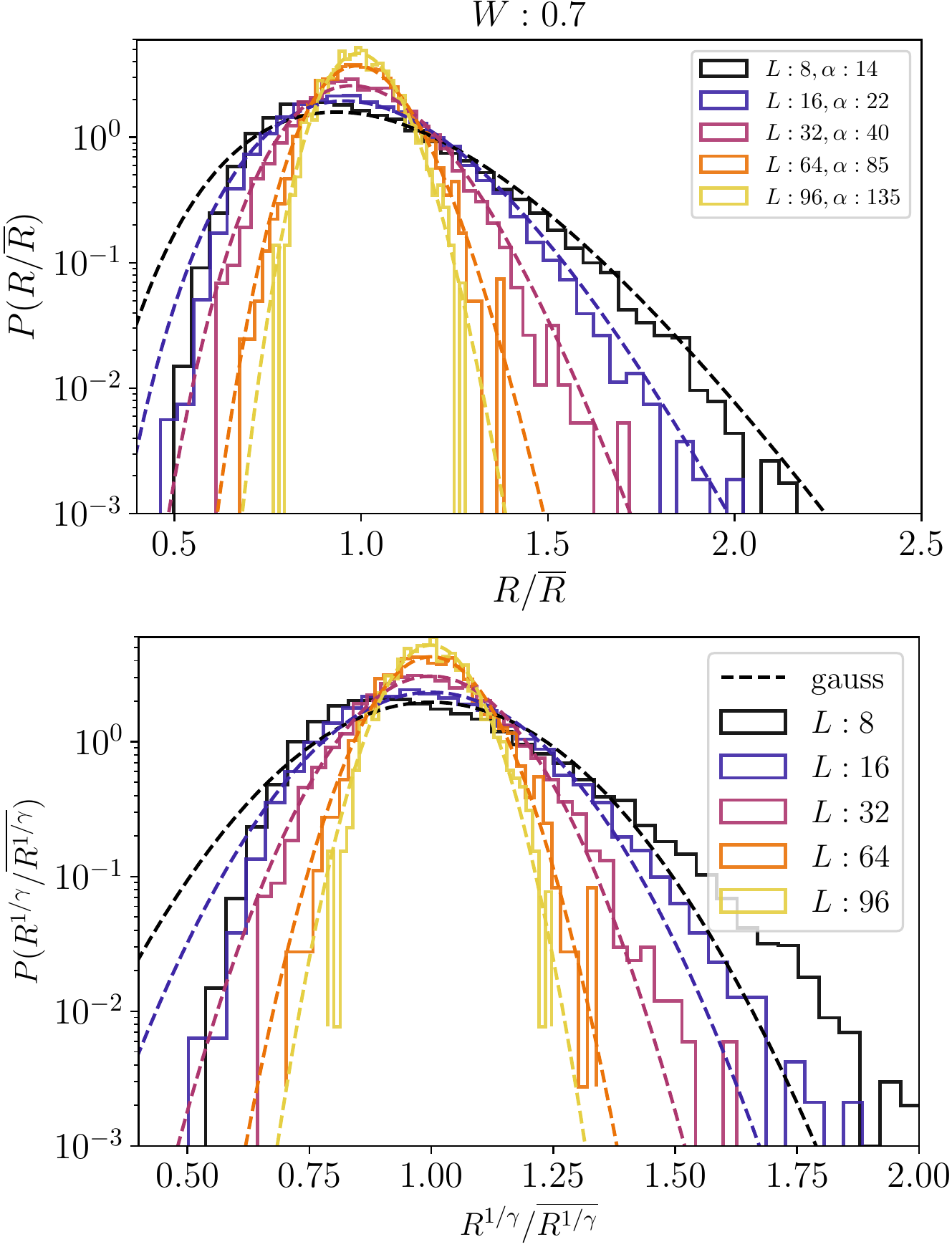}
    \caption{The top panel shows the probability distribution of the scaled resistance $R/\overline{R}$ for different system sizes at a disorder strength of $W = 0.7$ (weak subdiffusion).  The colored dashed lines denote an exponentially decaying function with parameter $\alpha$ (as described in the main text). The bottom panel displays the probability distribution $p(R^{1/\gamma}/\overline{R^{1/\gamma}})$, a quantity that we believe could be Gaussian-distributed in the TDL. The colored dashed lines are a fit to a Gaussian with the same standard deviation.}
    \label{fig:P_R_subdiff07}
\end{figure}

\begin{figure}
    \centering
    \includegraphics[width=\columnwidth]{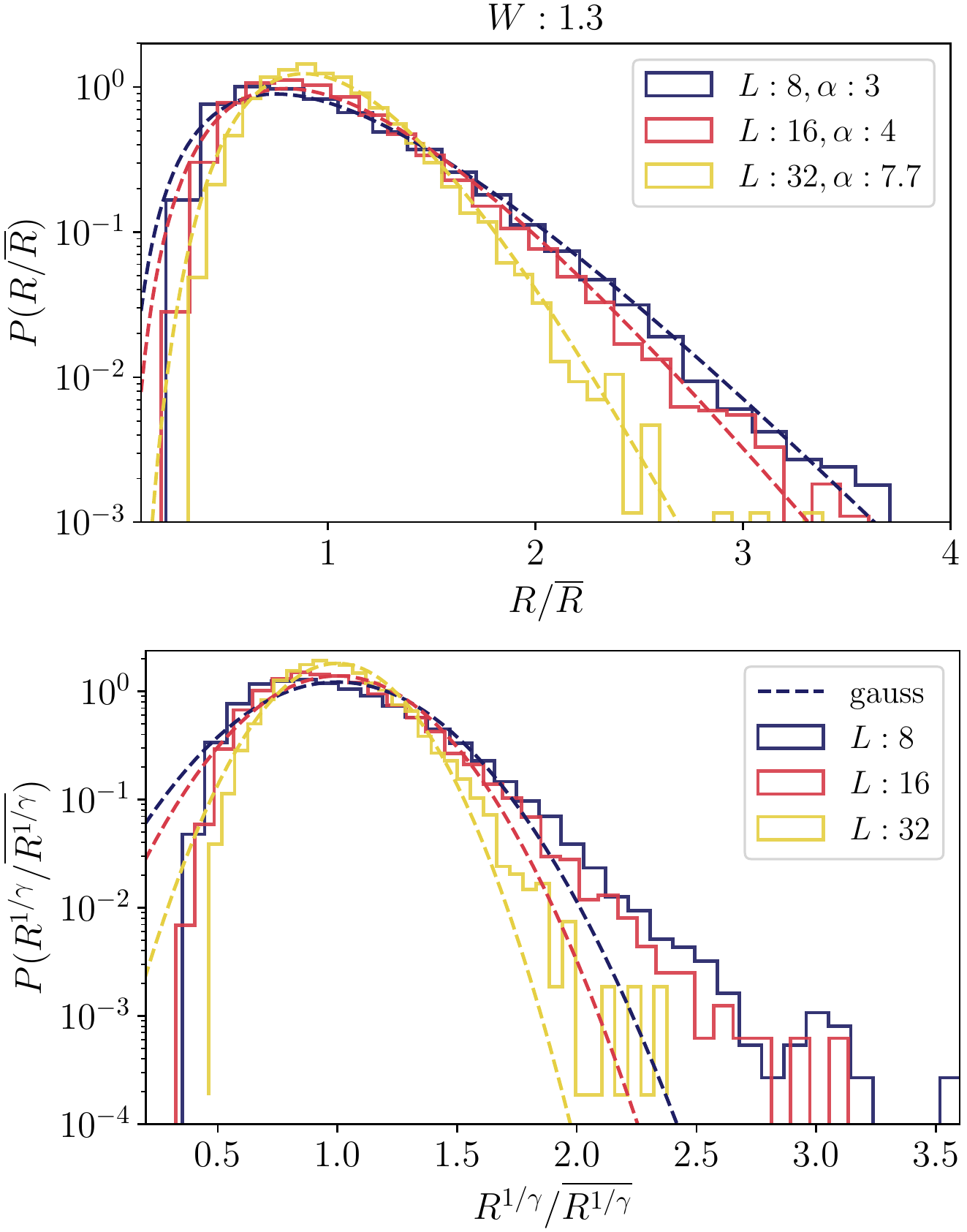}
    \caption{The top panel shows the probability distribution of the scaled resistance $R/\overline{R}$ for different system sizes at a disorder strength of $W = 1.3$ (stronger subdiffusion).  The colored dashed lines denote an exponentially decaying function with parameter $\alpha$ (as described in the main text). The bottom panel displays the probability distribution $p(R^{1/\gamma}/\overline{R^{1/\gamma}})$, a quantity that we believe could be Gaussian-distributed in the TDL. The colored dashed lines are a fit to a Gaussian with the same standard deviation.}
    \label{fig:P_R_subdiff13}
\end{figure}

With this rigorous analysis of the convergence properties of the NESS simulation at hand, we aim to make a reasonable statement about the occurrence of fat tails in the distribution of resistances. 
In Fig.~\ref{fig:R_R_diff} we show that at the boundary of the diffusion-subdiffusion transition, the distribution is indeed Gaussian and the additivity argument of Sec.~\ref{sec:additivity} holds therefore well. At fixed system size $L$, increasing the disorder $W$ further, there may be some form of tails developing. However, increasing $L$ at fixed $W$ the tails consistently decrease and the data can be fit with a function that decreases faster than a power-law with $\beta = 2$. In Fig.~\ref{fig:P_R_subdiff07} and Fig.~\ref{fig:P_R_subdiff13} we show the probability distributions of the resistance $R$ scaled by $\overline{R}$ for different system sizes for $W = 0.7$ and $W = 1.3$ respectively. We find that this data fits the gamma distribution $f(x) = \frac{(1+\alpha)^{(1+\alpha)}}{\Gamma(1+\alpha)} x^{\alpha}\exp{\left[-(1+\alpha)x\right]}$ well, indicating the possibility of exponentially decaying tails. We observe that $\alpha$ is an approximately linear function of system size for all $W$. Alternatively, we may also fit the data to a Levy-stable distribution which has a power-law decay, but the power of the decay turns out very large, far bigger than the two which would signal a transition to `fat tails', and furthermore the power increases with system size.

From the data, it is very hard to distinguish between a Levy-stable distribution with a large power (not displayed here) and the displayed exponential decay, as one needs both very precise numerics as well as large sample sizes to do so. For our numerics we used up to 40,000 samples, but exact simulations on very small systems suggest that in order to confidently discriminate the difference, larger samples might be needed. In addition to the above analysis we also test for the possibility that the simple quantity of $R^{1/\gamma}$ is additive and thus becomes Gaussian in the TDL. 
In the bottom panels of Fig.~\ref{fig:P_R_subdiff07} and Fig.~\ref{fig:P_R_subdiff13} we therefore show the probability distributions $p(R^{1/\gamma}/\overline{R^{1/\gamma}})$ and fit it to a Gaussian with the same standard-deviation. This certainly does not work well for small systems sizes, but as system sizes increase, the quality of the fit improves.

While the difference between a power-law decay with large power and an exponential decay cannot be resolved, we may rule out the existence of fat tails, given that the power of the Levy-stable distribution is significantly larger than 2 even for $W=1.3$, and furthermore it grows with increasing system size. While the precision due to finite $\chi$ (e.g. $\Delta R$) is not very high at larger $W$, and the choice of extrapolation can affect the precise values of $R$, in our view it cannot account for the lack of fat tails ($\beta \approx 2$) as that would require a significant redistribution of weight to the tails, see e.g. top Fig.~\ref{fig:P_R_subdiff13}. 
It remains imaginable that in order to capture the implications of rare regions we would require exponentially many Schmidt values for an accurate depiction of long-range coherence. However, to the best of our knowledge this would be outside the reach of any current numerical technique.

\section{Conclusion and perspective}
We have studied the distribution of resistances for a set of spin chains in the sub-diffusive regime. After validating our method on non-interacting models which show subdiffusion or localization, we proceed with the analysis of the Heisenberg model with random fields, which has a diffusive, a subdiffusive and an MBL region. The distribution of resistances does not show signs of long, power-law tails at the largest values of disorder we can study, a necessary ingredient of the phenomenology of rare regions (Griffiths physics) at the basis of many works on the ETH-to-MBL transition \cite{Gopalakrishnan2019dynamics}. In particular, we observe regular distributions decreasing fast in $R$, with variance shrinking as the system size increases, converging towards a self-averaging scenario opposite to the Griffiths one. A simple distribution compatible in the thermodynamic limit with all our data for non-interacting subdiffusive and localized, as well as for the interacting subdiffusive XXX model, is a Gaussian distribution of $R^{1/\gamma}$. We offer no microscopic mechanism to explain our observations, but we hope our work will provide the basis for a less speculative analysis of the Physics of the MBL-to-ETH transition.

\section*{Acknowledgements}
The authors thank Roderich Moessner for insightful discussions and Rajat K.\ Panda for discussions and collaboration on a related subject. AS, MS, and SRT are supported by the Trieste Institute for the Theory of Quantum Technologies. MS is supported by a Google Faculty Award. MZ acknowledges Grants No.~J1-7279, J1-1698 and program No. P1-0402 of the Slovenian Research Agency.

\bibliography{MBLbib}

\begin{thebibliography}{61}%
\makeatletter
\providecommand \@ifxundefined [1]{%
 \@ifx{#1\undefined}
}%
\providecommand \@ifnum [1]{%
 \ifnum #1\expandafter \@firstoftwo
 \else \expandafter \@secondoftwo
 \fi
}%
\providecommand \@ifx [1]{%
 \ifx #1\expandafter \@firstoftwo
 \else \expandafter \@secondoftwo
 \fi
}%
\providecommand \natexlab [1]{#1}%
\providecommand \enquote  [1]{``#1''}%
\providecommand \bibnamefont  [1]{#1}%
\providecommand \bibfnamefont [1]{#1}%
\providecommand \citenamefont [1]{#1}%
\providecommand \href@noop [0]{\@secondoftwo}%
\providecommand \href [0]{\begingroup \@sanitize@url \@href}%
\providecommand \@href[1]{\@@startlink{#1}\@@href}%
\providecommand \@@href[1]{\endgroup#1\@@endlink}%
\providecommand \@sanitize@url [0]{\catcode `\\12\catcode `\$12\catcode
  `\&12\catcode `\#12\catcode `\^12\catcode `\_12\catcode `\%12\relax}%
\providecommand \@@startlink[1]{}%
\providecommand \@@endlink[0]{}%
\providecommand \url  [0]{\begingroup\@sanitize@url \@url }%
\providecommand \@url [1]{\endgroup\@href {#1}{\urlprefix }}%
\providecommand \urlprefix  [0]{URL }%
\providecommand \Eprint [0]{\href }%
\providecommand \doibase [0]{http://dx.doi.org/}%
\providecommand \selectlanguage [0]{\@gobble}%
\providecommand \bibinfo  [0]{\@secondoftwo}%
\providecommand \bibfield  [0]{\@secondoftwo}%
\providecommand \translation [1]{[#1]}%
\providecommand \BibitemOpen [0]{}%
\providecommand \bibitemStop [0]{}%
\providecommand \bibitemNoStop [0]{.\EOS\space}%
\providecommand \EOS [0]{\spacefactor3000\relax}%
\providecommand \BibitemShut  [1]{\csname bibitem#1\endcsname}%
\let\auto@bib@innerbib\@empty
\bibitem [{\citenamefont {D'Alessio}\ \emph {et~al.}(2016)\citenamefont
  {D'Alessio}, \citenamefont {Kafri}, \citenamefont {Polkovnikov},\ and\
  \citenamefont {Rigol}}]{d2016quantum}%
  \BibitemOpen
  \bibfield  {author} {\bibinfo {author} {\bibfnamefont {L.}~\bibnamefont
  {D'Alessio}}, \bibinfo {author} {\bibfnamefont {Y.}~\bibnamefont {Kafri}},
  \bibinfo {author} {\bibfnamefont {A.}~\bibnamefont {Polkovnikov}}, \ and\
  \bibinfo {author} {\bibfnamefont {M.}~\bibnamefont {Rigol}},\ }\href@noop {}
  {\bibfield  {journal} {\bibinfo  {journal} {Advances in Physics}\ }\textbf
  {\bibinfo {volume} {65}},\ \bibinfo {pages} {239} (\bibinfo {year}
  {2016})}\BibitemShut {NoStop}%
\bibitem [{\citenamefont {Zotos}\ \emph {et~al.}(1997)\citenamefont {Zotos},
  \citenamefont {Naef},\ and\ \citenamefont {Prelovsek}}]{Zotos1997Transport}%
  \BibitemOpen
  \bibfield  {author} {\bibinfo {author} {\bibfnamefont {X.}~\bibnamefont
  {Zotos}}, \bibinfo {author} {\bibfnamefont {F.}~\bibnamefont {Naef}}, \ and\
  \bibinfo {author} {\bibfnamefont {P.}~\bibnamefont {Prelovsek}},\ }\href
  {\doibase 10.1103/PhysRevB.55.11029} {\bibfield  {journal} {\bibinfo
  {journal} {Phys. Rev. B}\ }\textbf {\bibinfo {volume} {55}},\ \bibinfo
  {pages} {11029} (\bibinfo {year} {1997})}\BibitemShut {NoStop}%
\bibitem [{\citenamefont {Anderson}(1958)}]{Anderson1958}%
  \BibitemOpen
  \bibfield  {author} {\bibinfo {author} {\bibfnamefont {P.}~\bibnamefont
  {Anderson}},\ }\href {\doibase 10.1103/PhysRev.109.1492} {\bibfield
  {journal} {\bibinfo  {journal} {Phys. Rev.}\ }\textbf {\bibinfo {volume}
  {109}},\ \bibinfo {pages} {1492} (\bibinfo {year} {1958})}\BibitemShut
  {NoStop}%
\bibitem [{\citenamefont {Basko}\ \emph {et~al.}(2006)\citenamefont {Basko},
  \citenamefont {Aleiner},\ and\ \citenamefont {Altshuler}}]{Basko:2006hh}%
  \BibitemOpen
  \bibfield  {author} {\bibinfo {author} {\bibfnamefont {D.~M.}\ \bibnamefont
  {Basko}}, \bibinfo {author} {\bibfnamefont {I.~L.}\ \bibnamefont {Aleiner}},
  \ and\ \bibinfo {author} {\bibfnamefont {B.~L.}\ \bibnamefont {Altshuler}},\
  }\href {http://www.sciencedirect.com/science/article/pii/S0003491605002630}
  {\bibfield  {journal} {\bibinfo  {journal} {Ann. Phys.}\ }\textbf {\bibinfo
  {volume} {321}},\ \bibinfo {pages} {1126} (\bibinfo {year}
  {2006})}\BibitemShut {NoStop}%
\bibitem [{\citenamefont {Gornyi}\ \emph {et~al.}(2005)\citenamefont {Gornyi},
  \citenamefont {Mirlin},\ and\ \citenamefont
  {Polyakov}}]{gornyi2005interacting}%
  \BibitemOpen
  \bibfield  {author} {\bibinfo {author} {\bibfnamefont {I.}~\bibnamefont
  {Gornyi}}, \bibinfo {author} {\bibfnamefont {A.}~\bibnamefont {Mirlin}}, \
  and\ \bibinfo {author} {\bibfnamefont {D.}~\bibnamefont {Polyakov}},\
  }\href@noop {} {\bibfield  {journal} {\bibinfo  {journal} {Phys. Rev. Lett.}\
  }\textbf {\bibinfo {volume} {95}},\ \bibinfo {pages} {206603} (\bibinfo
  {year} {2005})}\BibitemShut {NoStop}%
\bibitem [{\citenamefont {Nandkishore}\ and\ \citenamefont
  {Huse}(2015)}]{huse2015review}%
  \BibitemOpen
  \bibfield  {author} {\bibinfo {author} {\bibfnamefont {R.}~\bibnamefont
  {Nandkishore}}\ and\ \bibinfo {author} {\bibfnamefont {D.~A.}\ \bibnamefont
  {Huse}},\ }\href@noop {} {\bibfield  {journal} {\bibinfo  {journal} {Annual
  Review of Condensed Matter Physics}\ }\textbf {\bibinfo {volume} {6}},\
  \bibinfo {pages} {15} (\bibinfo {year} {2015})}\BibitemShut {NoStop}%
\bibitem [{\citenamefont {Abanin}\ and\ \citenamefont
  {Papi{\'c}}(2017)}]{abanin2017recent}%
  \BibitemOpen
  \bibfield  {author} {\bibinfo {author} {\bibfnamefont {D.~A.}\ \bibnamefont
  {Abanin}}\ and\ \bibinfo {author} {\bibfnamefont {Z.}~\bibnamefont
  {Papi{\'c}}},\ }\href@noop {} {\bibfield  {journal} {\bibinfo  {journal}
  {Annalen der Physik}\ }\textbf {\bibinfo {volume} {529}},\ \bibinfo {pages}
  {1700169} (\bibinfo {year} {2017})}\BibitemShut {NoStop}%
\bibitem [{\citenamefont {Imbrie}\ \emph {et~al.}(2017)\citenamefont {Imbrie},
  \citenamefont {Ros},\ and\ \citenamefont {Scardicchio}}]{imbrie2017review}%
  \BibitemOpen
  \bibfield  {author} {\bibinfo {author} {\bibfnamefont {J.~Z.}\ \bibnamefont
  {Imbrie}}, \bibinfo {author} {\bibfnamefont {V.}~\bibnamefont {Ros}}, \ and\
  \bibinfo {author} {\bibfnamefont {A.}~\bibnamefont {Scardicchio}},\ }\href
  {http://onlinelibrary.wiley.com/doi/10.1002/andp.201600278/epdf} {\bibfield
  {journal} {\bibinfo  {journal} {Annalen der Physik}\ }\textbf {\bibinfo
  {volume} {529}} (\bibinfo {year} {2017})}\BibitemShut {NoStop}%
\bibitem [{\citenamefont {Alet}\ and\ \citenamefont {N.}(2018)}]{Alet18}%
  \BibitemOpen
  \bibfield  {author} {\bibinfo {author} {\bibfnamefont {F.}~\bibnamefont
  {Alet}}\ and\ \bibinfo {author} {\bibfnamefont {L.}~\bibnamefont {N.}},\
  }\href@noop {} {\bibfield  {journal} {\bibinfo  {journal} {Comptes Rendus
  Physique}\ }\textbf {\bibinfo {volume} {19}},\ \bibinfo {pages} {498}
  (\bibinfo {year} {2018})}\BibitemShut {NoStop}%
\bibitem [{\citenamefont {Abanin}\ \emph {et~al.}(2019)\citenamefont {Abanin},
  \citenamefont {Altman}, \citenamefont {Bloch},\ and\ \citenamefont
  {Serbyn}}]{Abanin2019colloqium}%
  \BibitemOpen
  \bibfield  {author} {\bibinfo {author} {\bibfnamefont {D.~A.}\ \bibnamefont
  {Abanin}}, \bibinfo {author} {\bibfnamefont {E.}~\bibnamefont {Altman}},
  \bibinfo {author} {\bibfnamefont {I.}~\bibnamefont {Bloch}}, \ and\ \bibinfo
  {author} {\bibfnamefont {M.}~\bibnamefont {Serbyn}},\ }\href {\doibase
  10.1103/RevModPhys.91.021001} {\bibfield  {journal} {\bibinfo  {journal}
  {Rev. Mod. Phys.}\ }\textbf {\bibinfo {volume} {91}},\ \bibinfo {pages}
  {021001} (\bibinfo {year} {2019})}\BibitemShut {NoStop}%
\bibitem [{\citenamefont {Chandran}\ \emph {et~al.}(2016)\citenamefont
  {Chandran}, \citenamefont {Pal}, \citenamefont {Laumann},\ and\ \citenamefont
  {Scardicchio}}]{chandran2016many}%
  \BibitemOpen
  \bibfield  {author} {\bibinfo {author} {\bibfnamefont {A.}~\bibnamefont
  {Chandran}}, \bibinfo {author} {\bibfnamefont {A.}~\bibnamefont {Pal}},
  \bibinfo {author} {\bibfnamefont {C.}~\bibnamefont {Laumann}}, \ and\
  \bibinfo {author} {\bibfnamefont {A.}~\bibnamefont {Scardicchio}},\
  }\href@noop {} {\bibfield  {journal} {\bibinfo  {journal} {Physical Review
  B}\ }\textbf {\bibinfo {volume} {94}},\ \bibinfo {pages} {144203} (\bibinfo
  {year} {2016})}\BibitemShut {NoStop}%
\bibitem [{\citenamefont {De~Roeck}\ and\ \citenamefont
  {Huveneers}(2017)}]{de2017stability}%
  \BibitemOpen
  \bibfield  {author} {\bibinfo {author} {\bibfnamefont {W.}~\bibnamefont
  {De~Roeck}}\ and\ \bibinfo {author} {\bibfnamefont {F.}~\bibnamefont
  {Huveneers}},\ }\href@noop {} {\bibfield  {journal} {\bibinfo  {journal}
  {Physical Review B}\ }\textbf {\bibinfo {volume} {95}},\ \bibinfo {pages}
  {155129} (\bibinfo {year} {2017})}\BibitemShut {NoStop}%
\bibitem [{\citenamefont {Dorfman}(1999)}]{dorfman99}%
  \BibitemOpen
  \bibfield  {author} {\bibinfo {author} {\bibfnamefont {J.~R.}\ \bibnamefont
  {Dorfman}},\ }\href@noop {} {\emph {\bibinfo {title} {{An Introduction to
  Chaos in Nonequilibirum Statistical Mechanics}}}}\ (\bibinfo  {publisher}
  {Cambridge University Press},\ \bibinfo {year} {1999})\BibitemShut {NoStop}%
\bibitem [{\citenamefont {{Bar Lev}}\ \emph {et~al.}(2015)\citenamefont {{Bar
  Lev}}, \citenamefont {Cohen},\ and\ \citenamefont
  {Reichman}}]{Reichman2014Absence}%
  \BibitemOpen
  \bibfield  {author} {\bibinfo {author} {\bibfnamefont {Y.}~\bibnamefont {{Bar
  Lev}}}, \bibinfo {author} {\bibfnamefont {G.}~\bibnamefont {Cohen}}, \ and\
  \bibinfo {author} {\bibfnamefont {D.~R.}\ \bibnamefont {Reichman}},\
  }\href@noop {} {\bibfield  {journal} {\bibinfo  {journal} {Phys. Rev. Lett.}\
  }\textbf {\bibinfo {volume} {114}},\ \bibinfo {pages} {100601} (\bibinfo
  {year} {2015})}\BibitemShut {NoStop}%
\bibitem [{\citenamefont {Gopalakrishnan}\ \emph {et~al.}(2015)\citenamefont
  {Gopalakrishnan}, \citenamefont {M\"uller}, \citenamefont {Khemani},
  \citenamefont {Knap}, \citenamefont {Demler},\ and\ \citenamefont
  {Huse}}]{Gopalakrishnan2015low}%
  \BibitemOpen
  \bibfield  {author} {\bibinfo {author} {\bibfnamefont {S.}~\bibnamefont
  {Gopalakrishnan}}, \bibinfo {author} {\bibfnamefont {M.}~\bibnamefont
  {M\"uller}}, \bibinfo {author} {\bibfnamefont {V.}~\bibnamefont {Khemani}},
  \bibinfo {author} {\bibfnamefont {M.}~\bibnamefont {Knap}}, \bibinfo {author}
  {\bibfnamefont {E.}~\bibnamefont {Demler}}, \ and\ \bibinfo {author}
  {\bibfnamefont {D.~A.}\ \bibnamefont {Huse}},\ }\href {\doibase
  10.1103/PhysRevB.92.104202} {\bibfield  {journal} {\bibinfo  {journal} {Phys.
  Rev. B}\ }\textbf {\bibinfo {volume} {92}},\ \bibinfo {pages} {104202}
  (\bibinfo {year} {2015})}\BibitemShut {NoStop}%
\bibitem [{\citenamefont {Torres-Herrera}\ and\ \citenamefont
  {Santos}(2015)}]{torresherrera2015dynamics}%
  \BibitemOpen
  \bibfield  {author} {\bibinfo {author} {\bibfnamefont {E.~J.}\ \bibnamefont
  {Torres-Herrera}}\ and\ \bibinfo {author} {\bibfnamefont {L.~F.}\
  \bibnamefont {Santos}},\ }\href {\doibase 10.1103/PhysRevB.92.014208}
  {\bibfield  {journal} {\bibinfo  {journal} {Phys. Rev. B}\ }\textbf {\bibinfo
  {volume} {92}},\ \bibinfo {pages} {014208} (\bibinfo {year}
  {2015})}\BibitemShut {NoStop}%
\bibitem [{\citenamefont {Khait}\ \emph {et~al.}(2016)\citenamefont {Khait},
  \citenamefont {Gazit}, \citenamefont {Yao},\ and\ \citenamefont
  {Auerbach}}]{Khait2016spin}%
  \BibitemOpen
  \bibfield  {author} {\bibinfo {author} {\bibfnamefont {I.}~\bibnamefont
  {Khait}}, \bibinfo {author} {\bibfnamefont {S.}~\bibnamefont {Gazit}},
  \bibinfo {author} {\bibfnamefont {N.~Y.}\ \bibnamefont {Yao}}, \ and\
  \bibinfo {author} {\bibfnamefont {A.}~\bibnamefont {Auerbach}},\ }\href
  {\doibase 10.1103/PhysRevB.93.224205} {\bibfield  {journal} {\bibinfo
  {journal} {Phys. Rev. B}\ }\textbf {\bibinfo {volume} {93}},\ \bibinfo
  {pages} {224205} (\bibinfo {year} {2016})}\BibitemShut {NoStop}%
\bibitem [{\citenamefont {Luitz}\ \emph {et~al.}(2016)\citenamefont {Luitz},
  \citenamefont {Laflorencie},\ and\ \citenamefont {Alet}}]{Luitz2016Extended}%
  \BibitemOpen
  \bibfield  {author} {\bibinfo {author} {\bibfnamefont {D.~J.}\ \bibnamefont
  {Luitz}}, \bibinfo {author} {\bibfnamefont {N.}~\bibnamefont {Laflorencie}},
  \ and\ \bibinfo {author} {\bibfnamefont {F.}~\bibnamefont {Alet}},\ }\href
  {\doibase 10.1103/PhysRevB.93.060201} {\bibfield  {journal} {\bibinfo
  {journal} {Phys. Rev. B}\ }\textbf {\bibinfo {volume} {93}},\ \bibinfo
  {pages} {060201} (\bibinfo {year} {2016})}\BibitemShut {NoStop}%
\bibitem [{\citenamefont {Prelov\ifmmode~\check{s}\else \v{s}\fi{}ek}\ and\
  \citenamefont {Herbrych}(2017)}]{prelovchek2017self}%
  \BibitemOpen
  \bibfield  {author} {\bibinfo {author} {\bibfnamefont {P.}~\bibnamefont
  {Prelov\ifmmode~\check{s}\else \v{s}\fi{}ek}}\ and\ \bibinfo {author}
  {\bibfnamefont {J.}~\bibnamefont {Herbrych}},\ }\href {\doibase
  10.1103/PhysRevB.96.035130} {\bibfield  {journal} {\bibinfo  {journal} {Phys.
  Rev. B}\ }\textbf {\bibinfo {volume} {96}},\ \bibinfo {pages} {035130}
  (\bibinfo {year} {2017})}\BibitemShut {NoStop}%
\bibitem [{\citenamefont {Doggen}\ \emph {et~al.}(2018)\citenamefont {Doggen},
  \citenamefont {Schindler}, \citenamefont {Tikhonov}, \citenamefont {Mirlin},
  \citenamefont {Neupert}, \citenamefont {Polyakov},\ and\ \citenamefont
  {Gornyi}}]{doggen2018many}%
  \BibitemOpen
  \bibfield  {author} {\bibinfo {author} {\bibfnamefont {E.~V.~H.}\
  \bibnamefont {Doggen}}, \bibinfo {author} {\bibfnamefont {F.}~\bibnamefont
  {Schindler}}, \bibinfo {author} {\bibfnamefont {K.~S.}\ \bibnamefont
  {Tikhonov}}, \bibinfo {author} {\bibfnamefont {A.~D.}\ \bibnamefont
  {Mirlin}}, \bibinfo {author} {\bibfnamefont {T.}~\bibnamefont {Neupert}},
  \bibinfo {author} {\bibfnamefont {D.~G.}\ \bibnamefont {Polyakov}}, \ and\
  \bibinfo {author} {\bibfnamefont {I.~V.}\ \bibnamefont {Gornyi}},\ }\href
  {\doibase 10.1103/PhysRevB.98.174202} {\bibfield  {journal} {\bibinfo
  {journal} {Phys. Rev. B}\ }\textbf {\bibinfo {volume} {98}},\ \bibinfo
  {pages} {174202} (\bibinfo {year} {2018})}\BibitemShut {NoStop}%
\bibitem [{\citenamefont {Agarwal}\ \emph {et~al.}(2015)\citenamefont
  {Agarwal}, \citenamefont {Gopalakrishnan}, \citenamefont {Knap},
  \citenamefont {M{\"u}ller},\ and\ \citenamefont
  {Demler}}]{AgarwalAnomalousDiffusion}%
  \BibitemOpen
  \bibfield  {author} {\bibinfo {author} {\bibfnamefont {K.}~\bibnamefont
  {Agarwal}}, \bibinfo {author} {\bibfnamefont {S.}~\bibnamefont
  {Gopalakrishnan}}, \bibinfo {author} {\bibfnamefont {M.}~\bibnamefont
  {Knap}}, \bibinfo {author} {\bibfnamefont {M.}~\bibnamefont {M{\"u}ller}}, \
  and\ \bibinfo {author} {\bibfnamefont {E.}~\bibnamefont {Demler}},\
  }\href@noop {} {\bibfield  {journal} {\bibinfo  {journal} {Phys. Rev. Lett.}\
  }\textbf {\bibinfo {volume} {114}},\ \bibinfo {pages} {160401} (\bibinfo
  {year} {2015})}\BibitemShut {NoStop}%
\bibitem [{\citenamefont {\v{Z}nidari\v{c}}\ \emph {et~al.}(2016)\citenamefont
  {\v{Z}nidari\v{c}}, \citenamefont {Scardicchio},\ and\ \citenamefont
  {Varma}}]{Znidaric2016Diffusive}%
  \BibitemOpen
  \bibfield  {author} {\bibinfo {author} {\bibfnamefont {M.}~\bibnamefont
  {\v{Z}nidari\v{c}}}, \bibinfo {author} {\bibfnamefont {A.}~\bibnamefont
  {Scardicchio}}, \ and\ \bibinfo {author} {\bibfnamefont {V.~K.}\ \bibnamefont
  {Varma}},\ }\href@noop {} {\bibfield  {journal} {\bibinfo  {journal} {Phys.
  Rev. Lett.}\ }\textbf {\bibinfo {volume} {117}},\ \bibinfo {pages} {040601}
  (\bibinfo {year} {2016})}\BibitemShut {NoStop}%
\bibitem [{\citenamefont {Varma}\ \emph
  {et~al.}(2017{\natexlab{a}})\citenamefont {Varma}, \citenamefont {Lerose},
  \citenamefont {Pietracaprina}, \citenamefont {Goold},\ and\ \citenamefont
  {Scardicchio}}]{Varma2017Energy}%
  \BibitemOpen
  \bibfield  {author} {\bibinfo {author} {\bibfnamefont {V.~K.}\ \bibnamefont
  {Varma}}, \bibinfo {author} {\bibfnamefont {A.}~\bibnamefont {Lerose}},
  \bibinfo {author} {\bibfnamefont {F.}~\bibnamefont {Pietracaprina}}, \bibinfo
  {author} {\bibfnamefont {J.}~\bibnamefont {Goold}}, \ and\ \bibinfo {author}
  {\bibfnamefont {A.}~\bibnamefont {Scardicchio}},\ }\href {\doibase
  10.1088/1742-5468/aa668b} {\bibfield  {journal} {\bibinfo  {journal} {Journal
  of Statistical Mechanics: Theory and Experiment}\ }\textbf {\bibinfo {volume}
  {2017}},\ \bibinfo {pages} {053101} (\bibinfo {year}
  {2017}{\natexlab{a}})}\BibitemShut {NoStop}%
\bibitem [{\citenamefont {Mendoza-Arenas}\ \emph {et~al.}(2019)\citenamefont
  {Mendoza-Arenas}, \citenamefont {\ifmmode \check{Z}\else
  \v{Z}\fi{}nidari\ifmmode~\check{c}\else \v{c}\fi{}}, \citenamefont {Varma},
  \citenamefont {Goold}, \citenamefont {Clark},\ and\ \citenamefont
  {Scardicchio}}]{mendoza2018asymmetry}%
  \BibitemOpen
  \bibfield  {author} {\bibinfo {author} {\bibfnamefont {J.~J.}\ \bibnamefont
  {Mendoza-Arenas}}, \bibinfo {author} {\bibfnamefont {M.}~\bibnamefont
  {\ifmmode \check{Z}\else \v{Z}\fi{}nidari\ifmmode~\check{c}\else
  \v{c}\fi{}}}, \bibinfo {author} {\bibfnamefont {V.~K.}\ \bibnamefont
  {Varma}}, \bibinfo {author} {\bibfnamefont {J.}~\bibnamefont {Goold}},
  \bibinfo {author} {\bibfnamefont {S.~R.}\ \bibnamefont {Clark}}, \ and\
  \bibinfo {author} {\bibfnamefont {A.}~\bibnamefont {Scardicchio}},\ }\href
  {\doibase 10.1103/PhysRevB.99.094435} {\bibfield  {journal} {\bibinfo
  {journal} {Phys. Rev. B}\ }\textbf {\bibinfo {volume} {99}},\ \bibinfo
  {pages} {094435} (\bibinfo {year} {2019})}\BibitemShut {NoStop}%
\bibitem [{\citenamefont {Schulz}\ \emph {et~al.}(2018)\citenamefont {Schulz},
  \citenamefont {Taylor}, \citenamefont {Hooley},\ and\ \citenamefont
  {Scardicchio}}]{schulz2018energy}%
  \BibitemOpen
  \bibfield  {author} {\bibinfo {author} {\bibfnamefont {M.}~\bibnamefont
  {Schulz}}, \bibinfo {author} {\bibfnamefont {S.~R.}\ \bibnamefont {Taylor}},
  \bibinfo {author} {\bibfnamefont {C.~A.}\ \bibnamefont {Hooley}}, \ and\
  \bibinfo {author} {\bibfnamefont {A.}~\bibnamefont {Scardicchio}},\ }\href
  {\doibase 10.1103/PhysRevB.98.180201} {\bibfield  {journal} {\bibinfo
  {journal} {Phys. Rev. B}\ }\textbf {\bibinfo {volume} {98}},\ \bibinfo
  {pages} {180201} (\bibinfo {year} {2018})}\BibitemShut {NoStop}%
\bibitem [{\citenamefont {Karahalios}\ \emph {et~al.}(2009)\citenamefont
  {Karahalios}, \citenamefont {Metavitsiadis}, \citenamefont {Zotos},
  \citenamefont {Gorczyca},\ and\ \citenamefont {Prelov\ifmmode~\check{s}\else
  \v{s}\fi{}ek}}]{Karahalios2009finite}%
  \BibitemOpen
  \bibfield  {author} {\bibinfo {author} {\bibfnamefont {A.}~\bibnamefont
  {Karahalios}}, \bibinfo {author} {\bibfnamefont {A.}~\bibnamefont
  {Metavitsiadis}}, \bibinfo {author} {\bibfnamefont {X.}~\bibnamefont
  {Zotos}}, \bibinfo {author} {\bibfnamefont {A.}~\bibnamefont {Gorczyca}}, \
  and\ \bibinfo {author} {\bibfnamefont {P.}~\bibnamefont
  {Prelov\ifmmode~\check{s}\else \v{s}\fi{}ek}},\ }\href {\doibase
  10.1103/PhysRevB.79.024425} {\bibfield  {journal} {\bibinfo  {journal} {Phys.
  Rev. B}\ }\textbf {\bibinfo {volume} {79}},\ \bibinfo {pages} {024425}
  (\bibinfo {year} {2009})}\BibitemShut {NoStop}%
\bibitem [{\citenamefont {Steinigeweg}\ \emph {et~al.}(2016)\citenamefont
  {Steinigeweg}, \citenamefont {Herbrych}, \citenamefont {Pollmann},\ and\
  \citenamefont {Brenig}}]{Steinigeweg2016typicality}%
  \BibitemOpen
  \bibfield  {author} {\bibinfo {author} {\bibfnamefont {R.}~\bibnamefont
  {Steinigeweg}}, \bibinfo {author} {\bibfnamefont {J.}~\bibnamefont
  {Herbrych}}, \bibinfo {author} {\bibfnamefont {F.}~\bibnamefont {Pollmann}},
  \ and\ \bibinfo {author} {\bibfnamefont {W.}~\bibnamefont {Brenig}},\ }\href
  {\doibase 10.1103/PhysRevB.94.180401} {\bibfield  {journal} {\bibinfo
  {journal} {Phys. Rev. B}\ }\textbf {\bibinfo {volume} {94}},\ \bibinfo
  {pages} {180401} (\bibinfo {year} {2016})}\BibitemShut {NoStop}%
\bibitem [{\citenamefont {De~Roeck}\ \emph {et~al.}()\citenamefont {De~Roeck},
  \citenamefont {Hueveneers},\ and\ \citenamefont {Olla}}]{Roek2019prep}%
  \BibitemOpen
  \bibfield  {author} {\bibinfo {author} {\bibfnamefont {W.}~\bibnamefont
  {De~Roeck}}, \bibinfo {author} {\bibfnamefont {F.}~\bibnamefont
  {Hueveneers}}, \ and\ \bibinfo {author} {\bibfnamefont {S.}~\bibnamefont
  {Olla}},\ }\href@noop {} {\bibinfo  {journal} {arXiv:1909.07322}\
  }\BibitemShut {NoStop}%
\bibitem [{\citenamefont {Luitz}\ and\ \citenamefont
  {Bar~Lev}(2016)}]{Luitz2016Anomalous}%
  \BibitemOpen
\bibfield  {journal} {  }\bibfield  {author} {\bibinfo {author} {\bibfnamefont
  {D.~J.}\ \bibnamefont {Luitz}}\ and\ \bibinfo {author} {\bibfnamefont
  {Y.}~\bibnamefont {Bar~Lev}},\ }\href {\doibase
  10.1103/PhysRevLett.117.170404} {\bibfield  {journal} {\bibinfo  {journal}
  {Phys. Rev. Lett.}\ }\textbf {\bibinfo {volume} {117}},\ \bibinfo {pages}
  {170404} (\bibinfo {year} {2016})}\BibitemShut {NoStop}%
\bibitem [{\citenamefont {Griffiths}(1969)}]{Griffiths}%
  \BibitemOpen
  \bibfield  {author} {\bibinfo {author} {\bibfnamefont {R.~B.}\ \bibnamefont
  {Griffiths}},\ }\href {\doibase 10.1103/PhysRevLett.23.17} {\bibfield
  {journal} {\bibinfo  {journal} {Phys. Rev. Lett.}\ }\textbf {\bibinfo
  {volume} {23}},\ \bibinfo {pages} {17} (\bibinfo {year} {1969})}\BibitemShut
  {NoStop}%
\bibitem [{\citenamefont {Gopalakrishnan}\ \emph {et~al.}(2016)\citenamefont
  {Gopalakrishnan}, \citenamefont {Agarwal}, \citenamefont {Demler},
  \citenamefont {Huse},\ and\ \citenamefont
  {Knap}}]{Gopalakrishnan2016Griffiths}%
  \BibitemOpen
  \bibfield  {author} {\bibinfo {author} {\bibfnamefont {S.}~\bibnamefont
  {Gopalakrishnan}}, \bibinfo {author} {\bibfnamefont {K.}~\bibnamefont
  {Agarwal}}, \bibinfo {author} {\bibfnamefont {E.~A.}\ \bibnamefont {Demler}},
  \bibinfo {author} {\bibfnamefont {D.~A.}\ \bibnamefont {Huse}}, \ and\
  \bibinfo {author} {\bibfnamefont {M.}~\bibnamefont {Knap}},\ }\href {\doibase
  10.1103/PhysRevB.93.134206} {\bibfield  {journal} {\bibinfo  {journal} {Phys.
  Rev. B}\ }\textbf {\bibinfo {volume} {93}},\ \bibinfo {pages} {134206}
  (\bibinfo {year} {2016})}\BibitemShut {NoStop}%
\bibitem [{\citenamefont {Agarwal}\ \emph {et~al.}()\citenamefont {Agarwal},
  \citenamefont {Altman}, \citenamefont {Demler}, \citenamefont
  {Gopalakrishnan}, \citenamefont {Huse},\ and\ \citenamefont
  {Knap}}]{Agarwal2017rare}%
  \BibitemOpen
  \bibfield  {author} {\bibinfo {author} {\bibfnamefont {K.}~\bibnamefont
  {Agarwal}}, \bibinfo {author} {\bibfnamefont {E.}~\bibnamefont {Altman}},
  \bibinfo {author} {\bibfnamefont {E.}~\bibnamefont {Demler}}, \bibinfo
  {author} {\bibfnamefont {S.}~\bibnamefont {Gopalakrishnan}}, \bibinfo
  {author} {\bibfnamefont {D.~A.}\ \bibnamefont {Huse}}, \ and\ \bibinfo
  {author} {\bibfnamefont {M.}~\bibnamefont {Knap}},\ }\href {\doibase
  10.1002/andp.201600326} {\bibfield  {journal} {\bibinfo  {journal} {Annalen
  der Physik}\ }\textbf {\bibinfo {volume} {529}},\ \bibinfo {pages}
  {1600326}}\BibitemShut {NoStop}%
\bibitem [{\citenamefont {Potter}\ \emph {et~al.}(2015)\citenamefont {Potter},
  \citenamefont {Vasseur},\ and\ \citenamefont
  {Parameswaran}}]{Potter2015Universal}%
  \BibitemOpen
  \bibfield  {author} {\bibinfo {author} {\bibfnamefont {A.~C.}\ \bibnamefont
  {Potter}}, \bibinfo {author} {\bibfnamefont {R.}~\bibnamefont {Vasseur}}, \
  and\ \bibinfo {author} {\bibfnamefont {S.~A.}\ \bibnamefont {Parameswaran}},\
  }\href@noop {} {\bibfield  {journal} {\bibinfo  {journal} {Phys. Rev. X}\
  }\textbf {\bibinfo {volume} {5}},\ \bibinfo {pages} {031033} (\bibinfo {year}
  {2015})}\BibitemShut {NoStop}%
\bibitem [{\citenamefont {Vosk}\ \emph {et~al.}(2015)\citenamefont {Vosk},
  \citenamefont {Huse},\ and\ \citenamefont {Altman}}]{AltmanTheory2015}%
  \BibitemOpen
  \bibfield  {author} {\bibinfo {author} {\bibfnamefont {R.}~\bibnamefont
  {Vosk}}, \bibinfo {author} {\bibfnamefont {D.~A.}\ \bibnamefont {Huse}}, \
  and\ \bibinfo {author} {\bibfnamefont {E.}~\bibnamefont {Altman}},\
  }\href@noop {} {\bibfield  {journal} {\bibinfo  {journal} {Phys. Rev. X}\
  }\textbf {\bibinfo {volume} {5}},\ \bibinfo {pages} {031032} (\bibinfo {year}
  {2015})}\BibitemShut {NoStop}%
\bibitem [{\citenamefont {Gopalakrishnan}\ \emph {et~al.}(2017)\citenamefont
  {Gopalakrishnan}, \citenamefont {Islam},\ and\ \citenamefont
  {Knap}}]{Gopalakrishnan2017Noise}%
  \BibitemOpen
  \bibfield  {author} {\bibinfo {author} {\bibfnamefont {S.}~\bibnamefont
  {Gopalakrishnan}}, \bibinfo {author} {\bibfnamefont {K.~R.}\ \bibnamefont
  {Islam}}, \ and\ \bibinfo {author} {\bibfnamefont {M.}~\bibnamefont {Knap}},\
  }\href {\doibase 10.1103/PhysRevLett.119.046601} {\bibfield  {journal}
  {\bibinfo  {journal} {Phys. Rev. Lett.}\ }\textbf {\bibinfo {volume} {119}},\
  \bibinfo {pages} {046601} (\bibinfo {year} {2017})}\BibitemShut {NoStop}%
\bibitem [{\citenamefont {Vosk}\ and\ \citenamefont
  {Altman}(2013)}]{Vosk:2013kq}%
  \BibitemOpen
  \bibfield  {author} {\bibinfo {author} {\bibfnamefont {R.}~\bibnamefont
  {Vosk}}\ and\ \bibinfo {author} {\bibfnamefont {E.}~\bibnamefont {Altman}},\
  }\href@noop {} {\bibfield  {journal} {\bibinfo  {journal} {Phys. Rev. Lett.}\
  }\textbf {\bibinfo {volume} {110}},\ \bibinfo {pages} {067204} (\bibinfo
  {year} {2013})}\BibitemShut {NoStop}%
\bibitem [{\citenamefont {Gopalakrishnan}\ and\ \citenamefont
  {Nandkishore}(2014)}]{PhysRevB.90.224203}%
  \BibitemOpen
  \bibfield  {author} {\bibinfo {author} {\bibfnamefont {S.}~\bibnamefont
  {Gopalakrishnan}}\ and\ \bibinfo {author} {\bibfnamefont {R.}~\bibnamefont
  {Nandkishore}},\ }\href {\doibase 10.1103/PhysRevB.90.224203} {\bibfield
  {journal} {\bibinfo  {journal} {Phys. Rev. B}\ }\textbf {\bibinfo {volume}
  {90}},\ \bibinfo {pages} {224203} (\bibinfo {year} {2014})}\BibitemShut
  {NoStop}%
\bibitem [{\citenamefont {Anderson}\ \emph {et~al.}(1980)\citenamefont
  {Anderson}, \citenamefont {Thouless}, \citenamefont {Abrahams},\ and\
  \citenamefont {Fisher}}]{Anderson1980new}%
  \BibitemOpen
  \bibfield  {author} {\bibinfo {author} {\bibfnamefont {P.~W.}\ \bibnamefont
  {Anderson}}, \bibinfo {author} {\bibfnamefont {D.~J.}\ \bibnamefont
  {Thouless}}, \bibinfo {author} {\bibfnamefont {E.}~\bibnamefont {Abrahams}},
  \ and\ \bibinfo {author} {\bibfnamefont {D.~S.}\ \bibnamefont {Fisher}},\
  }\href {\doibase 10.1103/PhysRevB.22.3519} {\bibfield  {journal} {\bibinfo
  {journal} {Phys. Rev. B}\ }\textbf {\bibinfo {volume} {22}},\ \bibinfo
  {pages} {3519} (\bibinfo {year} {1980})}\BibitemShut {NoStop}%
\bibitem [{\citenamefont {Hiramoto}\ and\ \citenamefont
  {Abe}(1988)}]{Hisashi1988dynamics}%
  \BibitemOpen
  \bibfield  {author} {\bibinfo {author} {\bibfnamefont {H.}~\bibnamefont
  {Hiramoto}}\ and\ \bibinfo {author} {\bibfnamefont {S.}~\bibnamefont {Abe}},\
  }\href {\doibase 10.1143/JPSJ.57.230} {\bibfield  {journal} {\bibinfo
  {journal} {Journal of the Physical Society of Japan}\ }\textbf {\bibinfo
  {volume} {57}},\ \bibinfo {pages} {230} (\bibinfo {year} {1988})}\BibitemShut
  {NoStop}%
\bibitem [{\citenamefont {Pi{\'e}chon}(1996)}]{piechon1996anomalous}%
  \BibitemOpen
  \bibfield  {author} {\bibinfo {author} {\bibfnamefont {F.}~\bibnamefont
  {Pi{\'e}chon}},\ }\href@noop {} {\bibfield  {journal} {\bibinfo  {journal}
  {Physical review letters}\ }\textbf {\bibinfo {volume} {76}},\ \bibinfo
  {pages} {4372} (\bibinfo {year} {1996})}\BibitemShut {NoStop}%
\bibitem [{\citenamefont {Macé}\ \emph {et~al.}(2019)\citenamefont {Macé},
  \citenamefont {Laflorencie},\ and\ \citenamefont {Alet}}]{Mace2019many-body}%
  \BibitemOpen
  \bibfield  {author} {\bibinfo {author} {\bibfnamefont {N.}~\bibnamefont
  {Macé}}, \bibinfo {author} {\bibfnamefont {N.}~\bibnamefont {Laflorencie}},
  \ and\ \bibinfo {author} {\bibfnamefont {F.}~\bibnamefont {Alet}},\ }\href
  {\doibase 10.21468/SciPostPhys.6.4.050} {\bibfield  {journal} {\bibinfo
  {journal} {SciPost Phys.}\ }\textbf {\bibinfo {volume} {6}},\ \bibinfo
  {pages} {50} (\bibinfo {year} {2019})}\BibitemShut {NoStop}%
\bibitem [{\citenamefont {Varma}\ and\ \citenamefont {\ifmmode \check{Z}\else
  \v{Z}\fi{}nidari\ifmmode~\check{c}\else
  \v{c}\fi{}}(2019)}]{Varma2019Diffusive}%
  \BibitemOpen
  \bibfield  {author} {\bibinfo {author} {\bibfnamefont {V.~K.}\ \bibnamefont
  {Varma}}\ and\ \bibinfo {author} {\bibfnamefont {M.}~\bibnamefont {\ifmmode
  \check{Z}\else \v{Z}\fi{}nidari\ifmmode~\check{c}\else \v{c}\fi{}}},\ }\href
  {\doibase 10.1103/PhysRevB.100.085105} {\bibfield  {journal} {\bibinfo
  {journal} {Phys. Rev. B}\ }\textbf {\bibinfo {volume} {100}},\ \bibinfo
  {pages} {085105} (\bibinfo {year} {2019})}\BibitemShut {NoStop}%
\bibitem [{\citenamefont {\v{Z}nidari\v{c}}\ \emph {et~al.}(2008)\citenamefont
  {\v{Z}nidari\v{c}}, \citenamefont {Prosen},\ and\ \citenamefont
  {Prelov\v{s}ek}}]{vznidarivc2008many}%
  \BibitemOpen
  \bibfield  {author} {\bibinfo {author} {\bibfnamefont {M.}~\bibnamefont
  {\v{Z}nidari\v{c}}}, \bibinfo {author} {\bibfnamefont {T.}~\bibnamefont
  {Prosen}}, \ and\ \bibinfo {author} {\bibfnamefont {P.}~\bibnamefont
  {Prelov\v{s}ek}},\ }\href
  {https://journals.aps.org/prb/abstract/10.1103/PhysRevB.77.064426} {\bibfield
   {journal} {\bibinfo  {journal} {Phys. Rev. B}\ }\textbf {\bibinfo {volume}
  {77}},\ \bibinfo {pages} {064426} (\bibinfo {year} {2008})}\BibitemShut
  {NoStop}%
\bibitem [{\citenamefont {Pal}\ and\ \citenamefont {Huse}(2010)}]{pal2010many}%
  \BibitemOpen
  \bibfield  {author} {\bibinfo {author} {\bibfnamefont {A.}~\bibnamefont
  {Pal}}\ and\ \bibinfo {author} {\bibfnamefont {D.~A.}\ \bibnamefont {Huse}},\
  }\href@noop {} {\bibfield  {journal} {\bibinfo  {journal} {Physical review
  b}\ }\textbf {\bibinfo {volume} {82}},\ \bibinfo {pages} {174411} (\bibinfo
  {year} {2010})}\BibitemShut {NoStop}%
\bibitem [{\citenamefont {Luitz}\ \emph {et~al.}(2015)\citenamefont {Luitz},
  \citenamefont {Laflorencie},\ and\ \citenamefont {Alet}}]{alet2015}%
  \BibitemOpen
  \bibfield  {author} {\bibinfo {author} {\bibfnamefont {D.~J.}\ \bibnamefont
  {Luitz}}, \bibinfo {author} {\bibfnamefont {N.}~\bibnamefont {Laflorencie}},
  \ and\ \bibinfo {author} {\bibfnamefont {F.}~\bibnamefont {Alet}},\
  }\href@noop {} {\bibfield  {journal} {\bibinfo  {journal} {Phys. Rev. B}\
  }\textbf {\bibinfo {volume} {91}},\ \bibinfo {pages} {081103} (\bibinfo
  {year} {2015})}\BibitemShut {NoStop}%
\bibitem [{\citenamefont {Schollw{\"o}ck}(2011)}]{schollwock2011density}%
  \BibitemOpen
  \bibfield  {author} {\bibinfo {author} {\bibfnamefont {U.}~\bibnamefont
  {Schollw{\"o}ck}},\ }\href@noop {} {\bibfield  {journal} {\bibinfo  {journal}
  {Annals of Physics}\ }\textbf {\bibinfo {volume} {326}},\ \bibinfo {pages}
  {96} (\bibinfo {year} {2011})}\BibitemShut {NoStop}%
\bibitem [{\citenamefont {Mendoza-Arenas}\ \emph {et~al.}(2015)\citenamefont
  {Mendoza-Arenas}, \citenamefont {Clark},\ and\ \citenamefont
  {Jaksch}}]{Mendoza2015Coexistence}%
  \BibitemOpen
  \bibfield  {author} {\bibinfo {author} {\bibfnamefont {J.~J.}\ \bibnamefont
  {Mendoza-Arenas}}, \bibinfo {author} {\bibfnamefont {S.~R.}\ \bibnamefont
  {Clark}}, \ and\ \bibinfo {author} {\bibfnamefont {D.}~\bibnamefont
  {Jaksch}},\ }\href {\doibase 10.1103/PhysRevE.91.042129} {\bibfield
  {journal} {\bibinfo  {journal} {Phys. Rev. E}\ }\textbf {\bibinfo {volume}
  {91}},\ \bibinfo {pages} {042129} (\bibinfo {year} {2015})}\BibitemShut
  {NoStop}%
\bibitem [{\citenamefont {Gorini}\ \emph {et~al.}(1976)\citenamefont {Gorini},
  \citenamefont {Kossakowski},\ and\ \citenamefont
  {Sudarshan}}]{gorini1976completely}%
  \BibitemOpen
  \bibfield  {author} {\bibinfo {author} {\bibfnamefont {V.}~\bibnamefont
  {Gorini}}, \bibinfo {author} {\bibfnamefont {A.}~\bibnamefont {Kossakowski}},
  \ and\ \bibinfo {author} {\bibfnamefont {E.~C.~G.}\ \bibnamefont
  {Sudarshan}},\ }\href@noop {} {\bibfield  {journal} {\bibinfo  {journal}
  {Journal of Mathematical Physics}\ }\textbf {\bibinfo {volume} {17}},\
  \bibinfo {pages} {821} (\bibinfo {year} {1976})}\BibitemShut {NoStop}%
\bibitem [{\citenamefont {Lindblad}(1976)}]{lindblad1976generators}%
  \BibitemOpen
  \bibfield  {author} {\bibinfo {author} {\bibfnamefont {G.}~\bibnamefont
  {Lindblad}},\ }\href@noop {} {\bibfield  {journal} {\bibinfo  {journal}
  {Communications in Mathematical Physics}\ }\textbf {\bibinfo {volume} {48}},\
  \bibinfo {pages} {119} (\bibinfo {year} {1976})}\BibitemShut {NoStop}%
\bibitem [{\citenamefont {Breuer}\ and\ \citenamefont
  {Petruccione}(2002)}]{Breuer2002}%
  \BibitemOpen
  \bibfield  {author} {\bibinfo {author} {\bibfnamefont {H.-P.}\ \bibnamefont
  {Breuer}}\ and\ \bibinfo {author} {\bibfnamefont {F.}~\bibnamefont
  {Petruccione}},\ }\href@noop {} {\emph {\bibinfo {title} {The Theory of Open
  Quantum Systems}}}\ (\bibinfo  {publisher} {Oxford University Press},\
  \bibinfo {year} {2002})\BibitemShut {NoStop}%
\bibitem [{Note1()}]{Note1}%
  \BibitemOpen
  \bibinfo {note} {The temporal criterion is that the standard deviation of the
  average current over the previous 100 time-steps is less than $0.3\%$ of the
  average over the same period.}\BibitemShut {Stop}%
\bibitem [{\citenamefont
  {{\v{Z}}nidari{\v{c}}}(2019)}]{znidarivc2019nonequilibrium}%
  \BibitemOpen
  \bibfield  {author} {\bibinfo {author} {\bibfnamefont {M.}~\bibnamefont
  {{\v{Z}}nidari{\v{c}}}},\ }\href@noop {} {\bibfield  {journal} {\bibinfo
  {journal} {Physical Review B}\ }\textbf {\bibinfo {volume} {99}},\ \bibinfo
  {pages} {035143} (\bibinfo {year} {2019})}\BibitemShut {NoStop}%
\bibitem [{\citenamefont {Abrikosov}(1981)}]{abrikosov1981paradox}%
  \BibitemOpen
  \bibfield  {author} {\bibinfo {author} {\bibfnamefont {A.}~\bibnamefont
  {Abrikosov}},\ }\href@noop {} {\bibfield  {journal} {\bibinfo  {journal}
  {Solid State Communications}\ }\textbf {\bibinfo {volume} {37}},\ \bibinfo
  {pages} {997} (\bibinfo {year} {1981})}\BibitemShut {NoStop}%
\bibitem [{\citenamefont {Izrailev}\ \emph {et~al.}(1998)\citenamefont
  {Izrailev}, \citenamefont {Ruffo},\ and\ \citenamefont
  {Tessieri}}]{izrailev1998classical}%
  \BibitemOpen
  \bibfield  {author} {\bibinfo {author} {\bibfnamefont {F.}~\bibnamefont
  {Izrailev}}, \bibinfo {author} {\bibfnamefont {S.}~\bibnamefont {Ruffo}}, \
  and\ \bibinfo {author} {\bibfnamefont {L.}~\bibnamefont {Tessieri}},\
  }\href@noop {} {\bibfield  {journal} {\bibinfo  {journal} {Journal of physics
  A: Mathematical and general}\ }\textbf {\bibinfo {volume} {31}},\ \bibinfo
  {pages} {5263} (\bibinfo {year} {1998})}\BibitemShut {NoStop}%
\bibitem [{\citenamefont {Prosen}(2008)}]{prosen2008third}%
  \BibitemOpen
  \bibfield  {author} {\bibinfo {author} {\bibfnamefont {T.}~\bibnamefont
  {Prosen}},\ }\href@noop {} {\bibfield  {journal} {\bibinfo  {journal} {New
  Journal of Physics}\ }\textbf {\bibinfo {volume} {10}},\ \bibinfo {pages}
  {043026} (\bibinfo {year} {2008})}\BibitemShut {NoStop}%
\bibitem [{Note2()}]{Note2}%
  \BibitemOpen
  \bibinfo {note} {The odd-looking $\protect \sqrt {3}$ is here in order to
  have the same variance $W^2/3$ as one would have for the often-used box
  distribution $h_j \in [-W,W]$.}\BibitemShut {Stop}%
\bibitem [{\citenamefont {{\v{Z}}nidari{\v{c}}}\ and\ \citenamefont
  {Horvat}(2013)}]{znidarivc2013transport}%
  \BibitemOpen
  \bibfield  {author} {\bibinfo {author} {\bibfnamefont {M.}~\bibnamefont
  {{\v{Z}}nidari{\v{c}}}}\ and\ \bibinfo {author} {\bibfnamefont
  {M.}~\bibnamefont {Horvat}},\ }\href@noop {} {\bibfield  {journal} {\bibinfo
  {journal} {The European Physical Journal B}\ }\textbf {\bibinfo {volume}
  {86}},\ \bibinfo {pages} {67} (\bibinfo {year} {2013})}\BibitemShut {NoStop}%
\bibitem [{\citenamefont {Kohmoto}\ \emph {et~al.}(1983)\citenamefont
  {Kohmoto}, \citenamefont {Kadanoff},\ and\ \citenamefont
  {Tang}}]{kohmoto1983localization}%
  \BibitemOpen
  \bibfield  {author} {\bibinfo {author} {\bibfnamefont {M.}~\bibnamefont
  {Kohmoto}}, \bibinfo {author} {\bibfnamefont {L.~P.}\ \bibnamefont
  {Kadanoff}}, \ and\ \bibinfo {author} {\bibfnamefont {C.}~\bibnamefont
  {Tang}},\ }\href {\doibase 10.1103/PhysRevLett.50.1870} {\bibfield  {journal}
  {\bibinfo  {journal} {Phys. Rev. Lett.}\ }\textbf {\bibinfo {volume} {50}},\
  \bibinfo {pages} {1870} (\bibinfo {year} {1983})}\BibitemShut {NoStop}%
\bibitem [{\citenamefont {Ostlund}\ \emph {et~al.}(1983)\citenamefont
  {Ostlund}, \citenamefont {Pandit}, \citenamefont {Rand}, \citenamefont
  {Schellnhuber},\ and\ \citenamefont {Siggia}}]{ostlund1983one}%
  \BibitemOpen
  \bibfield  {author} {\bibinfo {author} {\bibfnamefont {S.}~\bibnamefont
  {Ostlund}}, \bibinfo {author} {\bibfnamefont {R.}~\bibnamefont {Pandit}},
  \bibinfo {author} {\bibfnamefont {D.}~\bibnamefont {Rand}}, \bibinfo {author}
  {\bibfnamefont {H.~J.}\ \bibnamefont {Schellnhuber}}, \ and\ \bibinfo
  {author} {\bibfnamefont {E.~D.}\ \bibnamefont {Siggia}},\ }\href {\doibase
  10.1103/PhysRevLett.50.1873} {\bibfield  {journal} {\bibinfo  {journal}
  {Phys. Rev. Lett.}\ }\textbf {\bibinfo {volume} {50}},\ \bibinfo {pages}
  {1873} (\bibinfo {year} {1983})}\BibitemShut {NoStop}%
\bibitem [{\citenamefont {Varma}\ \emph
  {et~al.}(2017{\natexlab{b}})\citenamefont {Varma}, \citenamefont
  {de~Mulatier},\ and\ \citenamefont {\ifmmode \check{Z}\else
  \v{Z}\fi{}nidari\ifmmode~\check{c}\else \v{c}\fi{}}}]{Varma2017fractality}%
  \BibitemOpen
  \bibfield  {author} {\bibinfo {author} {\bibfnamefont {V.~K.}\ \bibnamefont
  {Varma}}, \bibinfo {author} {\bibfnamefont {C.}~\bibnamefont {de~Mulatier}},
  \ and\ \bibinfo {author} {\bibfnamefont {M.}~\bibnamefont {\ifmmode
  \check{Z}\else \v{Z}\fi{}nidari\ifmmode~\check{c}\else \v{c}\fi{}}},\ }\href
  {\doibase 10.1103/PhysRevE.96.032130} {\bibfield  {journal} {\bibinfo
  {journal} {Phys. Rev. E}\ }\textbf {\bibinfo {volume} {96}},\ \bibinfo
  {pages} {032130} (\bibinfo {year} {2017}{\natexlab{b}})}\BibitemShut
  {NoStop}%
\bibitem [{\citenamefont {Gopalakrishnan}\ and\ \citenamefont
  {Parameswaran}(2019)}]{Gopalakrishnan2019dynamics}%
  \BibitemOpen
  \bibfield  {author} {\bibinfo {author} {\bibfnamefont {S.}~\bibnamefont
  {Gopalakrishnan}}\ and\ \bibinfo {author} {\bibfnamefont {S.}~\bibnamefont
  {Parameswaran}},\ }\href {https://arxiv.org/abs/1908.10435} {\bibfield
  {journal} {\bibinfo  {journal} {arXiv: 1908.10435}\ } (\bibinfo {year}
  {2019})}\BibitemShut {NoStop}%
\end{thebibliography}%

\end{document}